\documentclass[final,3p,times]{elsarticle}

\usepackage{amssymb}
\usepackage{tabularx}
\usepackage{array}
\usepackage{wrapfig}
\usepackage{makecell}
\usepackage{multirow}
\usepackage{comment}
\usepackage{mathtools}
\usepackage[bottom]{footmisc}

\DeclarePairedDelimiter\abs{\lvert}{\rvert}

\journal{Physica A}

\begin{document}

{\let\thefootnote\relax\footnote{2020.This manuscript version is made available under the CC-BY-NC-ND 4.0 license}} 

\begin{frontmatter}



\title{The use of scaling properties to detect relevant changes in financial time series: a new visual warning tool}

\author[ad1,ad5]{Ioannis P. Antoniades \thanks{Corresponding author. Email: iantoniades@auth.gr }}
\author[ad2]{Giuseppe Brandi}
\author[ad4]{L. G. Magafas}
\author[ad2,ad3]{T. Di Matteo}

\address[ad1]{Aristotle University of Thessaloniki, Physics Department, Thessaloniki, }
\address[ad2]{Department of Mathematics, King's College London, The Strand, London, WC2R 2LS, UK}
\address[ad3]{Complexity Science Hub Vienna, Josefstaedter Strasse 39, A 1080 Vienna, Austria}
\address[ad4]{International Hellenic University, Physics Department, Complex Systems Laboratory}
\address[ad5]{American College of Thessaloniki, Division of Science \& Technology}

\begin{abstract}
The dynamical evolution of multiscaling in financial time series is investigated using time-dependent Generalized Hurst Exponents (GHE), $H_q$, for various values of the parameter $q$. Using $H_q$, we introduce a new visual methodology to algorithmically detect critical changes in the scaling of the underlying complex time-series. The methodology involves the degree of multiscaling at a particular time instance, the multiscaling trend which is calculated by the Change-Point Analysis method, and a rigorous evaluation of the statistical significance of the results. Using this algorithm, we have identified particular patterns in the temporal co-evolution of the different $H_q$ time-series. These GHE patterns, distinguish in a statistically robust way, not only between time periods of uniscaling and multiscaling, but also among different types of multiscaling: symmetric multiscaling (M) and asymmetric multiscaling (A). Asymmetric multiscaling can also be robustly divided into three other subcategories. We apply the visual methodology to time-series comprising of daily close prices of four stock market indices: two major ones (S\&P~500 and Tokyo-NIKKEI) and two peripheral ones (Athens Stock Exchange general Index and Bombay-SENSEX). Results show that multiscaling varies greatly with time: time periods of strong multiscaling behavior and time periods of uniscaling behavior are interchanged while transitions from uniscaling to multiscaling behavior occur before critical market events, such as stock market bubbles. Moreover, particular asymmetric multiscaling patterns appear during critical stock market eras and provide useful information about market conditions. In particular, they can be used as 'fingerprints' of a turbulent market period as well as provide warning signals for an upcoming stock market 'bubble'. The applied visual methodology also appears to distinguish between exogenous and endogenous stock market crises, based on the observed patterns before the actual events. The visual methodology is sufficiently general to be applicable for the description of the dynamical evolution of multiscaling properties of any complex system.

\end{abstract}


\begin{highlights}
\item Visual warning tool for financial time series based on scaling analysis
\item Application of time-dependent Generalized Hurst Exponent method to financial timeseries 
\end{highlights}



\end{frontmatter}


\section{Introduction}
\label{intro}

The study of scaling in financial systems has been a field of investigation for many years now (\cite{Dima2007}, \cite{DimaAsteDaco2005}, \cite{DimaAsteDaco2003}, \cite{Mand1963}, \cite{CalvFish2002}, \cite{BoucPottMeye2000}, \cite{MantStan1995}, \cite{Leba2001}, \cite{Kaiz2001}, \cite{Scal1998}, \cite{BartMellDiMa2007}, \cite{LiuLuxDiMa2007}, \cite{LiuDiMaLux2008}, \cite{Mand1997}, \cite{MiloHatiBarn2020},\cite{barunik2010hurst},\cite{kristoufek2011multifractal},\cite{jiang2019multifractal},\cite{barunik2012understanding}). These studies have shown that financial series, especially from stock-markets, display multiscaling, which is nowadays  widely accepted as empirical stylized fact of financial time series. The (multi)scaling property of time series is particularly important in risk management, especially when the model used assumes independence of asset returns. In fact, the lack of this assumption to hold, may severely bias risk measures, especially if there is long-range dependence and this is acting with a different degree across the time series statistical moments. In recent years, multiscaling has been adopted as a formalism in two different branches of quantitative finance, i.e. econophysics and mathematical finance. The former devoted most of the attention to price and returns series in order to understand the source of multiscaling from an empirical and theoretical point of view \cite{mantegna_stanley_book,dacorogna_book,MantStan1995,Dima2007,CalvFish2002,lux1,lux_marchesi,DimaAsteDaco2005,buonocore2020interplay} and has recently identified a new stylized fact which relates (non-linearly) the strength of multiscaling and the dependence between stocks \cite{buonocore2020interplay}. The latter instead builds on the work of \cite{roughvola} on rough volatility and has been used to construct stochastic models with anti-persistent volatility dynamics \cite{roughvola,roughvola2,roughvola3,roughvola4}. Although these research fields try to answer different research questions, it is important to recognize the relevance that multiscaling has attained in finance. Multiscaling has been understood to originate from one or more phenomena related to trading dynamics. In particular, it can be attributed to (i) the fat tails in price change distributions, (ii) the auto-correlation of the absolute value of log-returns, (iii) liquidity dynamics, or (iv) (non-linear) correlation between high and low returns generated by the different time horizon of traders and the consequent volumes traded. It can also be caused by the endogeneity of markets for which a given order generates many other orders. The latter occurs especially in markets where algorithmic trading is prevalent \cite{brandi2020statistics}.
 However, scaling in a financial time series has also been shown to vary with time. For example, there have been studies trying to link this variation with dynamical elements in the underlying title such as, for instance, the level of stability of a firm (\cite{MoraDiMatteo2011}). In \cite{DrozKowa2018} and \cite{DrozOswi2015} the authors discuss, by using Multi-Fractal Detrended Fluctuation Analysis method (MF-DFA), the dynamical evolution of the $f(\alpha)$ \textit{vs.} $\alpha$ multifractal spectrum in financial and other types of time-series, not only in terms of its width $\Delta \alpha = \alpha_{max}-\alpha_{min}$ but also in terms of its 'asymmetry', \textit{i.e.} looking at the evolution of the shape (skewness) of the spectrum and relating it to market events and underling dynamics. Other studies have tried to associate a time-varying Hurst exponent as a measure of the dynamically changing scaling of a financial time-series, with the development of stock-market bubbles (\cite{YalaRossMcKe2011}, \cite{FernSancMuno2017} and references therein), trading signals (\cite{KrohSkou2018}) and predictability of an index \cite{CapoGila2017}, raising the question whether scaling analysis can be used as a signaling tool for financial markets (\cite{GrechMazu2004},\cite{GrechPamu2008},\cite{Mitr2012}).

 In the present study, we aim to contribute towards this discussion by studying the dynamical evolution of multiscaling using the structure function approach, also known as the Generalized Hurst exponent (GHE) method \citep{DimaAsteDaco2003,Dima2007,jiang2019multifractal}, on time-series from four stock market indices, two major ones, S\&P~500 and Tokyo-NIKKEI), and two from peripheral markets, Athens Stock Exchange General Index (ASE) and Bombay-SENSEX. Employing the GHE method, the generalized Hurst exponents, $H_q$, are calculated for various values of the parameter $q$ corresponding to time scaling of the $q$-moment of the series difference distribution for a time delay $\tau$. In the time-dependent GHE approach, time-series of $H_q$ are generated for a range of $q$ values, by partitioning the underlying time-series into (usually overlapping) time segments and calculating the $H_q$ values for each segment. Looking at the relative values of the $H_q$ for the various $q$ at a particular time segment, one can evaluate the degree of multiscaling during that period. Alternative methods to GHE can also be used to extract the scaling exponent from time series, such as Rescaled range (R/S) analysis (\cite{Hurs1951,jiang2019multifractal}), MF-DFA (\cite{KantZschKosc2002}) and the Wavelet Transform Modulus Maxima (WTMM) introduced by \citep{muzy1991wavelets,muzy1993}. A more complete discussion on the use and misuse of various Hurst exponent estimation methods is given by Serinaldi \cite{Seri2010}, suggesting caution on the method used depending on the type of time-series considered. Recently, \cite{buonocore2020interplay} showed that the results retrieved by GHE methodology and MF-DFA are qualitatively equivalent while \citep{barunik2010hurst} showed empirically that the GHE approach outperforms the other methods under different data specifications. For this reason, throughout this work we will use the GHE methodology.
 
 The scope of this work is threefold: (i) At a given time period, to detect \textit{differences} among the GHE temporal profiles of a time-series and the respective profiles of a surrogate randomly generated time-series of similar volatility temporal profile as the original series, using the exact same estimation method for both. (ii) To detect temporal changes in the GHE profiles of a time-series. We are thus interested in detecting statistically significant \textit{differences}, rather than absolute values, of GHE's relative to a specific reference series (the surrogate series) and the temporal evolution of these differences. (iii) to identify recurrent patterns in the temporal profiles that may correspond to particular market conditions. These patterns could characteristically emerge before or after critical time periods such as a stock-market bubble. In the first case, they can be used as warning signals for a particular future market event. In order to provide a rigorous definition of such patterns in GHE profiles and a systematic way to detect them, we introduce a visual methodology to algorithmically detect critical changes in the scaling of the underlying complex time-series. The methodology involves the strength of multiscaling at a particular time instance, the multiscaling trend, which is calculated by the Change-Point Analysis method, and a rigorous evaluation of the statistical significance of the identified patterns, by comparing to the output of the same analysis applied to randomly generated surrogate time-series that are constructed so that they have the same volatility temporal profile as the real series. Using this algorithm, we have identified particular patterns in the temporal co-evolution of the different GHE time-series. These patterns, that we call GHE Temporal Patterns (TP), distinguish in a statistically robust way, not only between time periods of uniscaling and multiscaling, but also among different types of multiscaling: \textit{symmetric} multiscaling and \textit{asymmetric} multiscaling. The later type is characterized by a time asymmetric dynamic of the scaling exponents for the extreme $q$ values, $q_1$ and $q_2$. The methodology shows that asymmetric multiscaling itself can be robustly divided into three subcategories that correspond to different dynamics. By applying the above visual methodology to historical data of the four indices mentioned above, we find that critical events are preceded by asymmetric multiscaling patterns thus highlighting a warning signal. We also find that such behaviour is in general stronger for endogenous crisis as the Dot.com bubble, the 1991 Japanese bubble, or the 2000 Athens bubble, but much weaker for exogenous generated ones, such as the 2008 global financial crisis. Furthermore, we discuss the physical connection of the multiscaling TP's to underlying market trading dynamics.
 
 The paper is structured as follows. Section \ref{methods} is devoted to the presentation of the methods and implementation used in the paper, section  \ref{results} shows results of an empirical application of the methodology to stock market indices, while sections \ref{discussion} and \ref{conclusions} are devoted to the discussion of the results, conclusions and future work for the further development of the method used in this study and its possible application to financial time series or time series of other complex systems.

\section{Description of methods}
\label{methods}
\subsection{Generalized Hurst Exponents}
\label{GHE}

The Hurst exponent (\cite{Hurs1951}, \cite{HursBlacSima1965}) is a well-known tool used to study the scaling behavior of time series coming from any dynamical process. To compute the scaling exponents, it is necessary to study the $q$-order moments of the absolute value of the increments of the stochastic process \cite{Dima2007}. In particular, the process $(X_t)$ with stationary increments is analysed through
 \begin{equation}\label{eq:Kq}
\Xi(\tau,q)= \mathbb{E}\left[|X_{( t+ \tau)} -X_t|^q\right]\sim K_q\tau^{qH_q},
 \end{equation}
 where $q=\{q_1,q_2,\dots,q_M\}$ is the set of evaluated moments, $\tau=\{\tau_1,\tau_2,\dots,\tau_N\}$ is the set of time aggregation used to compute the process increments, $K_q$ is the $q$-moment for $\tau=1$ and $H_q$ is the so called generalized Hurst exponent, which is a function of $q$. The function $q H_q$ is concave \cite{Mand1963,Mand1997} and codifies the scaling exponents of the process. A multiscaling proxy can be obtained by fitting the measured scaling exponents with a second degree polynomial (\cite{buonocore2020interplay,buonocore2016measuring}) of the form\footnote{Technical details of the choice of this functional form can be found in \cite{buonocore2020interplay,buonocore2016measuring}.}
 
 \begin{equation}\label{mult_proxy2}
 qH_q=Aq+Bq^2,
 \end{equation}
  or equivalently \cite{brandi2020statistics}:
   \begin{equation}\label{mult_proxy3}
    H_q=A+Bq,
 \end{equation}
  
where $A$ and $B$ are two constants. In this setting, the measured $B$, $\widehat B$, represents the curvature of $qH_q$. If $\widehat B=0$, $H_q$ does not depend on $q$, i.e. $H_q=H$ for all $q$, hence the process is uniscaling, while if $\widehat B\neq0$, the process is multiscaling \cite{Dima2007,brandi2020statistics,buonocore2020interplay,buonocore2016measuring}. For $q=1$, the GHE is equivalent to the original Hurst exponent. Notice also that for $q=2$, $\Xi(\tau,2)$ is proportional to the auto-correlation function of $X_t$. For $H_1=0.5$, the evolution of the system in state-space is equivalent to a random walk, i.e. the underlying process is purely stochastic (diffusive). For a single variable time series, this is equivalent to saying that at any given time, the value of the series is equally likely to go up as it is to go down. For $H_1>0.5$, the system evolves faster than stochastic diffusion (super-diffusive process), which implies that -for a single-variable series- if a change occurs in one direction (up or down), it is more likely that the next change will be in the same direction rather than in the opposite. In such a case, the underlying process is characterized as a {\it persistent} process. Finally, for $H_1< 0.5$, the system evolves slower than stochastic diffusion (sub-diffusive process). For a single-variable series, this implies that, if a change occurs in one direction (up or down), it is more likely that the next change will be in the opposite direction. In the latter case, the process is characterized as an {\it anti-persistent} process. For the calculation of GHE's of higher and more positive values of $q$, the largest differences in a series are weighted more than smaller differences in (\ref{eq:Kq}) and therefore large-$q$ GHE's emphasize the tails of the distribution of differences. Conversely, lower (less positive) values of $q$ weigh small differences more than large ones. Computing a broad spectrum of GHE's, for several spread-out values of $q$, provides a more detailed 'signature' of the underlying dynamics of the system compared to considering only the original Hurst exponent. However, using high values of $q$ can bias the results if the data analysed is characterised by distributions with fat tails. In particular, for $q>\alpha$, where $\alpha$ is the tail exponent of the distribution of the data, the $q$-moments are not well defined. This introduces a bias on the expected value which in turn, produces a bias in the GHE estimation. Since financial time series are generally fat tailed, the choice of $q$ is relevant.

\subsection{Weighted GHE's}
\label{wGHE}
Recently, Morales \& Di~Matteo in \cite{MoraDiMatteo2011} have proposed a modification of the GHE, the weighted GHE ({\it w}GHE), by modifying the way the time averaging is carried out in Equation (\ref{eq:Kq}). Specifically, while taking the sum within a time interval $[t-\Delta t, t]$ of length $\Delta t$, each term of the time series is weighed by a factor that depends on how far back from the present time $t$ the term lies: the farther in the past, the less this term is weighed, so that more recent times have a higher contribution to the calculation of the moments in Equation (\ref{eq:Kq}). Thus, the averaging in Equation (\ref{eq:Kq}) is replaced by the following definition: For any function $f$ of a dynamic variable $X_t$, we have

\begin{equation}
\label{eq:wAver}
\mathbb{E}\left[f \left( X_{t} \right)\right]_{ \theta }= \sum _{s=0}^{ \Delta t-1}w_{s} \left(  \theta  \right) f \left( X_{(t-s)} \right),
\end{equation}

where the weighting factor $w_{s}$ is an exponentially decaying function of time defined as:

\begin{equation}
\label{eq:Ws}
w_{s} \left(  \theta  \right) =w_{o} \left(  \theta  \right) e^{-\frac{s}{ \theta }},
\end{equation}

where $\theta$ is the characteristic time for which $w$ drops to $1/e$ and $w_{o}=w_{o} \left(  \theta  \right) =\frac{1-e^{-\frac{1}{ \theta}}}{1-e^{-\frac{ \Delta t}{\theta}}}$ is a normalization constant that ensures that the sum of all weights $w$ within the interval $\Delta t$ equals 1. Thus, Equation (\ref{eq:Kq}) is now replaced by its weighted sum equivalent:

\begin{equation}
\label{eq:wKq}
\Xi(\tau,q,\theta)= \mathbb{E}\left[|X_{( t+ \tau)} -X_t|^q\right]_{\theta}\sim K_q\tau^{qH_q^{(\theta)}},
\end{equation}

where $H_q^{(\theta)}$ is the {\it w}GHE corresponding to a characteristic time $\theta$. Throughout the rest of this paper we will use the {\it w}GHE version as defined above. Its main advantage is that it allows one to use a fixed window $\Delta t$ for all calculations, varying only the characteristic time $\theta$ in order to increase or decrease the weighting of the short-term past relative to the long-term past. This provides enough data to obtain accurate estimates for {\it w}GHE's and at the same time gives flexibility in setting the characteristic weighting time scale, thus adjusting smoothly the importance of recent past to distant past in GHE computation. 
 Besides {\it w}GHE's, we also estimated the time evolution of the volatility of an index, defined as the standard deviation of the weighted log returns over a time window equal to $\Delta t$. For volatility calculation, the averaging was again carried out as a weighted average with a characteristic time $\theta$ using Equation (\ref{eq:wAver}):

\begin{equation}
\label{eq:volatility}
V(t) = \sigma\left( log\left(\frac{X_{\tau+1}}{X_\tau}\right) w_{t} \left(  \theta  \right)\right)_{\Delta t},
\end{equation}

where $log(...)$ is the natural logarithm, $\sigma(...)_{\Delta t}$ denotes standard deviation of the  series for $\tau=1...\Delta t-1$ and the time weighting factor $w$ is given by Equation (\ref{eq:Ws}).

\subsection{Surrogate stock market indices}
\label{SurrIndex}
In order to make sure that our results are not numerical artefacts of the finite data sizes due to the relatively short time windows used in the {\it w}GHE computations, we applied the same calculations on \textit{surrogate} time series. In order to produce such series, we did not apply the 'shuffling' method often used for this purpose, according to which the surrogate is constructed by a random permutation of the original time-series percentage differences, in order to destroy any long-term correlations of the original data. Instead, for each market index studied, we created a respective surrogate index as follows: Starting at the actual close price of the particular index at an initial date (the first date for which data was available), closing prices of all subsequent dates were artificially generated by a 'random walk' procedure, in which the day-to-day log price change was picked from a normal distribution with mean equal to zero and variance equal to the weighted average volatility $V(t)$ of the actual index at that particular date $t$. In this way, the surrogate index day-to-day relative price changes are randomly chosen, but the volatility variation of the surrogate index (i.e. the average magnitude of the relative daily changes) matches the temporal volatility profile of the actual index.
The reason for making this choice is that, in the present study, the surrogate series merely serve as reference series for the purpose of subtracting the effect in multiscaling properties that are \textit{solely} due to the finite-sized (short length) data segments used in calculations from any {\it w}GHE temporal variations of the real index that are the cause of the underlying market dynamics. In other words, the surrogate index serves as a measure of the "noise level" for the {\it w}GHE of the real indices, which, after being subtracted, will enable a more accurate quantitative evaluation of the departure of observed multiscaling behavior from a randomly generated finite data set whose distribution of differences is normal, by construction. Randomly shuffling a real index, on the other hand, destroys any temporal correlations but maintains the precise distribution of changes intact. Therefore, comparisons of the {\it w}GHE temporal profiles of the real index with the respective profiles of a shuffled surrogate, does not seclude the effect of the non-normal character of real price distributions, an effect that we want to measure. Another obvious choice for a surrogate index would be a randomly generated index with price changes picked from a normal distribution of uniform variance in time (i.e. ignoring the effect of a time-varying market volatility). In the present study, we chose to include the effects of the volatility variation with time, in order to subtract any residual effect it may have on the {\it w}GHE's. In this way, we are sure to measure the effects on the {\it w}GHE's profiles coming from the departure of the price change distributions from being normal (although with time-varying variance), as well as any temporal correlations within the close price time series themselves.

\section{Results}
\label{results}
\subsection{Data description}
\label{DataDescr}
For our analysis, we have used 4 stock market indices: New York stock exchange index (S\&P~500), Tokyo stock exchange index (NIKKEI), Athens Stock Exchange general index (ASE) and the Bombay stock index (SENSEX). Table \ref{tab:tabdata} shows the time period in which the data is analysed and the number of trading days in each series.

\begin{table}[!h]
\centering
\begin{tabular}{|c|c|c|}
\hline
\textbf{Market} & \textbf{Time period} & \textbf{Trading days} \\ \hline
S\&P~500         & 1927-2020            & 23138                 \\ \hline
NIKKEI          & 1969-2020            & 13068                 \\ \hline
ASE             & 1991-2020            & 7146                  \\ \hline
SENSEX          & 2001-2020            & 5450                  \\ \hline
\end{tabular}
\caption{Time periods and the number of trading days analysed for each stock market.}
\label{tab:tabdata}
\end{table}
For each data series $X_t$, we used daily log prices, which is defined as the natural logarithm of the closing price of the index at each day, i.e. $X_t=log(P_t)$, where $P_t$ is the closing price of the index at time $t$.


\subsection{Standardized GHE, multiscaling proxies and parameter definition}
\label{Proxies}

We use a convenient normalization for $H_q$ defined as follows:
\begin{equation}
\label{eq:NormH}
H_q^{'(\theta)} =  \frac{H_q^{(\theta)}-H^{*}}{\sigma\left( H_q^{surr(\theta)} \right)},
\end{equation}

where $H^{*}$ is the value of Hurst exponent expected for a perfectly random series ($H^{*}=0.5$) and $\sigma\left( H^{surr(\theta)}_q \right)$ is the standard deviation of $H^{surr}_\theta(q)$, the {\it w}GHE of the respective surrogate series calculated over the entire timeline, which is computed as:

\begin{equation}
\label{eq:stdev}
\sigma\left(H^{surr}_q\right) = \sqrt{\frac{\sum_{\tau=1}^N \left(  H^{surr}_q(\tau)-\mathbb{E}\left[H^{surr}_q\right]\right)^2 }{N-1}},
\end{equation}

where $\mathbb{E}\left[H^{surr}_q\right]$ is the average of the series:

\begin{equation}
\label{eq:mean}
\mathbb{E}\left[ H^{surr}_q\right] = \frac{1}{N} \sum_{t=1}^N H^{surr}_q(t),
\end{equation}

and $N$ is the total number of points in the time-series. This type of normalized GHE (to which we will refer, from now on, as the 'standardized' GHE in order to distinguish it from usual normalized versions that contain the standard deviation of the real series itself) has a convenient interpretation: a value of $H_q^{'(\theta)}\approx0$ signifies an underlying time-series with the same behavior as a random series. For other values, $H_q^{'(\theta)}$ is equal to the number of standard deviations of $H_q^{surr(\theta)}$ that the real index $H_q^{(\theta)}$ is above 0.5. The standard deviation of $H_q^{surr(\theta)}$ is a measure of the variability of the $H_q$ series of a random index and thus conveniently measures the degree of 'noise level', i.e. the variability of any $H_q$ series that is due to finite data size effects and not to the actual underlying dynamics (apart from the dynamical changes in volatility which -in the present work- are included in the generation of the random surrogate). Therefore, division by the standard deviation of the surrogate series, enables a quantification of the statistical strength of the observed persistent, anti-persistent or uniscaling/multiscaling behavior of the real time-series at each time period, compared to a random signal for which the \textit{w}GHE's are computed with the same window size $\Delta t$ and characteristic weighting factor $\theta$.

In order to assess multiscaling, we use two alternative measures:
\begin{itemize}
    \item Multiscaling \textit{width} $W_{q_1,q_2}$
    \item Multiscaling \textit{curvature} (or \textit{depth}) $B$ of Equation \ref{mult_proxy3}.
\end{itemize}
The multiscaling width $W_{q_1,q_2}$ is computed as the difference between the $H_{q_1}$ and $H_{q_2}$, i.e. 

\begin{equation}
\label{eq:W}
W_{q_1,q_2}=H_{q_1} - H_{q_2},
\end{equation}

 and conveys information on the span of the $H_q$ parameter. Conversely, the multiscaling curvature $B$ is computed as the linear fit between $q$ and $H_q^{(\theta)}$, as described in Equation (\ref{mult_proxy3}) (\cite{brandi2020statistics,buonocore2020interplay}). If the process is uniscaling, both measures should be approximately zero as $H_q^{(\theta)}$ doesn't depend on $q$. In order to run our procedure, we have to specify some input parameters, i.e. $\tau$, $q$, and $\Delta t$ and $\theta$ in Equation \ref{eq:wKq}. Regarding the maximum $\tau$, we use 19 days, as prescribed in \cite{DimaAsteDaco2003,DimaAsteDaco2005}. Similarly to the standardized $H_q$, in order to detect statistically significant multiscaling at time $t$, the 'width' of the {\it w}GHE $q$-spectrum for extreme $q$ values, $q_1$ and $q_2$ in respect, is also standardized as

\begin{equation}
\label{eq:Wnorm}
W'_{q_1,q_2}(t) = \frac{W_{0.1,4}(t)}{\sigma\left( W^{surr}_{q_1,q_2} \right)},
\end{equation}

where $\sigma\left( W^{surr}_{q_1,q_2} \right)$ is the \textit{pooled} standard deviation of the difference between surrogate series $H^{surr}_{q_1}(t)$ and $H^{surr}_{q_2}(t)$ given by:

\begin{equation}
\label{eq:Stdpooled}
\sigma\left( W^{surr}_{q_1,q_2} \right) = \sqrt{\sigma\left( H^{surr}_{q_1} \right)^2+\sigma\left( H^{surr}_{q_2} \right)^2},
\end{equation}

Finally, in order to compute the series of $B(t)$, we use the series $H_q(t)$ for several values of $q$ within a range $q'_1-q'_2$. The number of $q$ values affects the accuracy of determining $B(t)$ by the least squares linear fit to $H_q$ \textit{vs.} $q$ data for each time $t$. A number of about 20 $q$ are adequate for a good quality fit, yielding 'p-values', on the average, above 0.98, and in the worst case (rare outliers) 0.85. Similar to $W'_{0.1,4}(t)$, we standardize $B$ by using the standard deviation of $B$ computed on the surrogate data, $\sigma\left( B^{surr} \right)$:

\begin{equation}
\label{eq:Bnorm}
B'(t)=\frac{B(t)}{\sigma\left( B^{surr} \right)}.
\end{equation}

$\sigma\left( B^{surr} \right)$ is calculated via Equation (\ref{eq:stdev}) replacing $H_q(\tau)$ by $B^{surr}(\tau)$.

 Regarding the choice of the extreme values of $q$, we used two sets: For the multiscaling width $W_{q_1,q_2}$, we used a large span, $q_1=0.1$ and $q_2=4$, in order to capture the strong 'biasing' effect of the tails of the price change distributions as it has been reported elsewhere \cite{buonocore2020interplay,brandi2020statistics} for financial time-series. We want to include this "biased" version of the width in order to capture the \textit{dynamics} of such bias and spot any transitions it may reveal in time. For the multiscaling proxy $B$, on the other hand, we used a short span $q'_1=0.1$ and $q'_2=1$ with a step of $\Delta q=0.04$ in order to concentrate on the small $q$ values that mostly weigh the small price changes and thus emphasize the center of the price change distributions. The step of $\Delta q=0.04$ provides 23 $H_q$ values for each point in time $t$, and thus the quality of the linear fit yielding $B$ is very good.

\subsubsection{Choice of $\Delta t$ and $\theta$}
\begin{figure}[h!]
\includegraphics[scale=0.3]{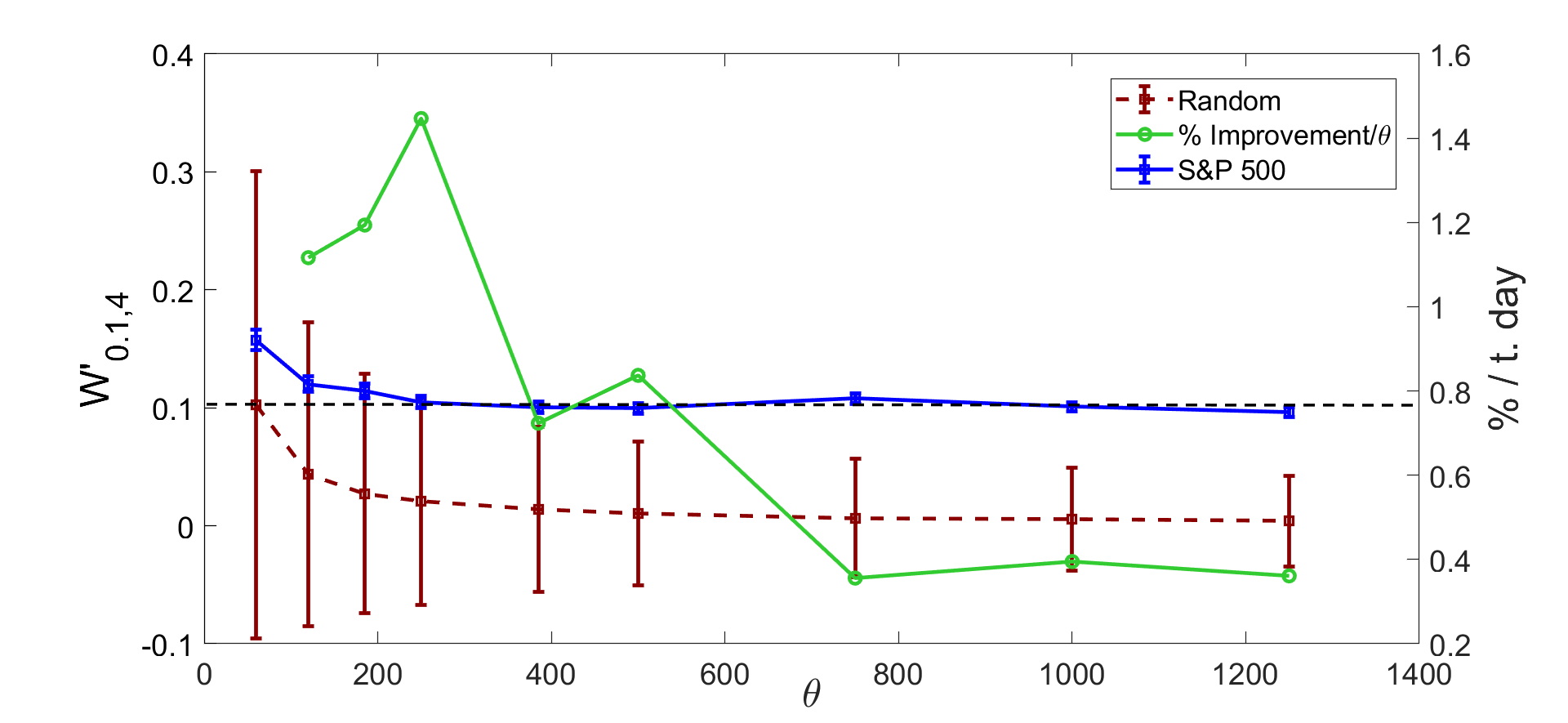}
\caption{\label{fig:OptimumTheta} Average width $W_{1,4}$ of real S\&P~500 and a random surrogate of the same data length as S\&P~500 for various values of $\Delta t$. For calculations $\Delta t = \theta$. Error bars for the real series correspond to the standard error of the mean. Error bars for the surrogate series correspond to the standard deviation of the surrogate series $W_{0.1,4}$. The horizontal line shows the plateau the of the S\&P~500 width. The \% rate of reduction of the deviation of the width from the plateau value per one day increase in $\Delta t$ is shown on the right axis.}
\end{figure}

One of the most important issues concerning the time-dependent {\it w}GHE's is the choice of the $\Delta t$ and $\theta$ parameters which represent the size of the time window and the time weighting parameter within that window that directly pertain to the {\it w}GHE's calculations. The optimum choice should be the result of a trade-off between reducing the finite-size effects (that increase the smaller $\Delta t$ and $\theta$ are), and capturing the short-term changes in multiscaling and {\it w}GHE's, a task for which the smaller $\Delta t$ and $\theta$, the better. If the time window length and time weighting parameter are too short, finite size effects will overwhelm the amount of multiscaling caused by the real dynamics. If, on the other hand, they are too large, finite size effects are ameliorated, but possible short-term multiscaling variations in the real dynamics are lost because they are averaged out in time. Moreover, the averaging-out effect may lead to another undesirable effect: to obtain spurious multiscaling estimation for the time period immediately after some extreme tail event which biases the width of the {\it w}GHE spectrum, especially for the large $q$ values. For example, if one picks $\Delta t=750$ trading days, then a large tail event will cause a bias in the {\it w}GHE's for a period of approximately 750 days (the characteristic decay time of the bias also depends on $\theta$). For a choice of $\Delta t=120$ trading days instead, the forward in time 'contamination' of the {\it w}GHE spectrum will have a much shorter duration, but finite-size effects will rise considerably for such small $\Delta t$. In order to make a proper choice, first of all we set $\Delta t=\theta$. This choice is arbitrary, but, without loss of generality, corresponds to a time window for which the last day in the past is weighted by a factor $1/e$ less than the most recent day. Then, $\Delta t$ is determined by the rule that it should be: i) as small as possible (in order to capture short-term dynamical changes and avoid long-term 'contamination' of multiscaling due to large tail events) and ii) sufficiently large that the noise level due to finite-size effects is satisfactorily low. In order to make a plausible choice meeting the above criteria, we calculated the width $W_{1,4}(t)$ time-series with a range of $\Delta t$ from 60-1250 trading days for the S\&P~500 index as well as its random surrogate. Then, for each $\Delta t$, we calculate the average value of the width of the time-series and plot it $vs.$ $\Delta t$ in figure~\ref{fig:OptimumTheta}. For the real index, error bars correspond to the standard \textit{error} of the average, whereas for the surrogate index the error bars correspond to the standard \textit{deviation} of the surrogate $W_{1,4}$ time-series. We see that the average width decreases with $\Delta t$, both for the real index and the surrogate, as finite-size effects are reduced as $\Delta t$ rises. For the real index, the average width naturally reaches a plateau that corresponds to the actual multiscaling strength (on the average) of the index, whereas for the random surrogate the width slowly drops to zero, the theoretical value for a random series. The dashed horizontal line in the figure shows the value of the plateau calculated as the average of the widths for $\Delta t=250,375,500,750,1000$ and $1250$. We see that already for $\Delta t \sim 250$ the finite size effects have considerably reduced and the value of the average width of the real index has reached the plateau value well within standard error. We also see that $\sim 250$ is the smallest value of $\Delta t$ for which the width of the real series is above at one standard deviation of the surrogate series average width, which means that for this value of $\Delta t$, the observed multiscaling is statistically strong (above the 'noise' level). Finally, for each of the depicted values of $\Delta t$, we plot the rate of \%~improvement of the average width of the real series per day, if $\Delta t$ is increased beyond each specific value shown. We observe that for the lowest values of $\Delta t$ the rate of improvement is high. Again, $\Delta t \sim 250$ is the smallest value for which this rate significantly drops, which means that if $\Delta t$ is increased beyond $\sim 250$ trading days, the improvement in noise level reduction is not significant. For all the above reasons we chose $\Delta t=\theta=250$ trading days as our optimum window size and time weighting factor.        

\subsection{{\it w}GHE's vs. time}
\label{GHEresults}
\begin{figure}[h!]
\begin{center}
\includegraphics[height=0.5\textheight,width=1.05\textwidth]{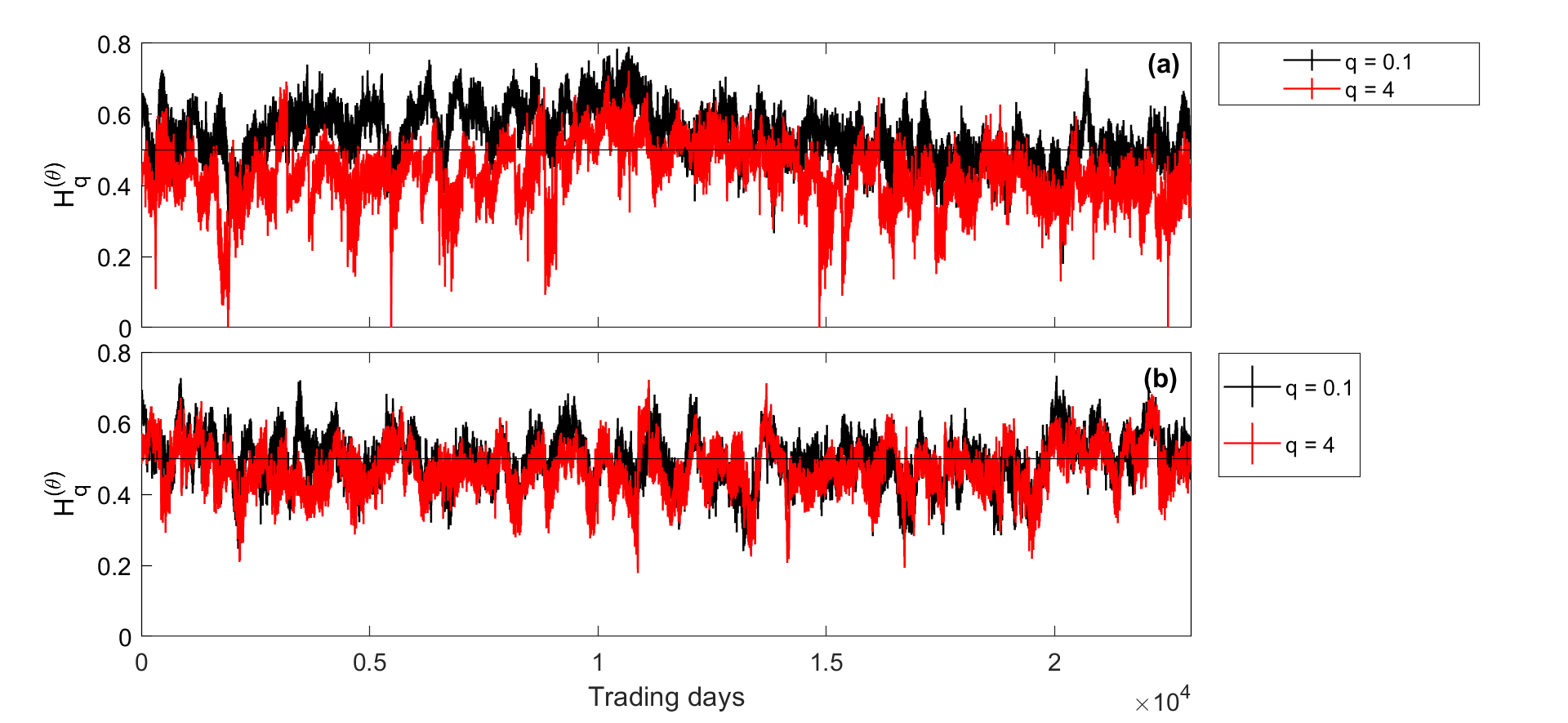} 
\caption{\label{fig:SP500_RAW_250} Time series of {\it w}GHE's for $q=0.1$ and $q=4$ and $\theta=250$ trading days of (a) the SP500 index log close prices and (b) the SP500 surrogate index log close prices. The width of each line is equal to two standard errors of $H_q$ as determined by the least squares fitting performed by the GHE algorithm in \cite{MoraDiMatteo2011}.}
\end{center}
\end{figure}

\begin{figure}[h!]
\begin{center}
\includegraphics[height=0.5\textheight,width=1.05\textwidth]{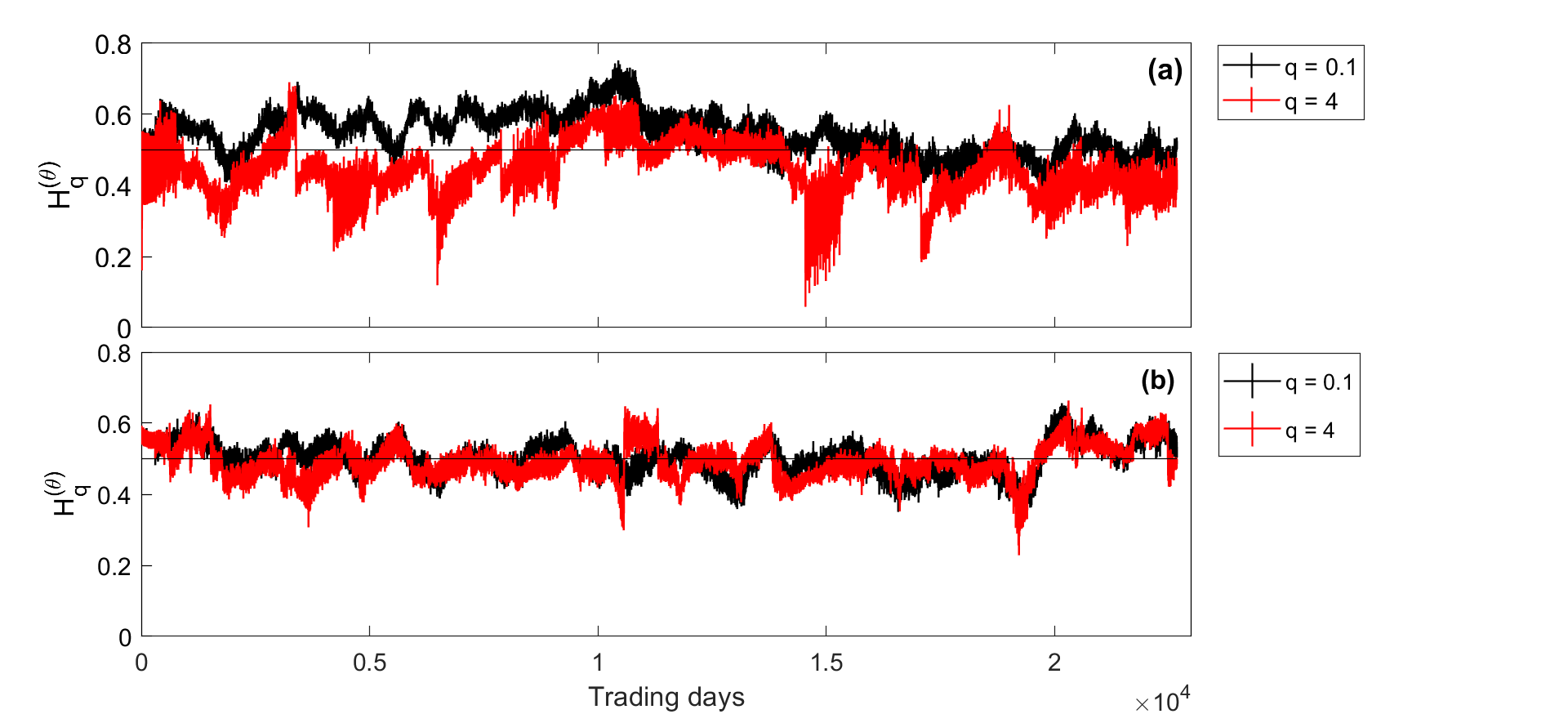} 
\caption{\label{fig:SP500_RAW_750} Time series of {\it w}GHE's for $q=0.1$ and $q=4$ and $\theta=250$ trading days of (a) the SP500 index log close prices and (b) the SP500 surrogate index log close prices.}
\end{center}
\end{figure}

Before we proceed with our main analysis of the GHE time-series and the introduction of the scaling pattern identification methodology, we present in figure~\ref{fig:SP500_RAW_250}, the raw (not-standardized) time series of $H_{0.1}^{(\theta)}$ and $H_4^{(\theta)}$ for the S\&P~500 index log prices and $\Delta t=\theta=250$ trading days together with the {\it w}GHE's of the respective S\&P~500 surrogate series. For comparison, in figure~\ref{fig:SP500_RAW_750}, we show the same quantities for $\Delta t=\theta=750$ trading days.\footnote{Throughout this paper, when we refer to the 'surrogate series' of a particular stock market index, we mean the randomly generated index according to the procedure highlighted in section \ref{SurrIndex}, where the surrogate series matches the temporal profile of the volatility of the real index.} The width of each line shown corresponds to the uncertainty of the {\it w}GHE's, which is equal to one standard deviation above and below the mean value of the {\it w}GHE's, as determined by the fitting procedure in the GHE algorithm. This error depends on finite-size effects and varies significantly for each time segment considered based on the quality of the least squares fit in the GHE algorithm for a particular time segment. The error is larger the smaller the values of $\Delta t$ (and $\theta$) and is also larger the bigger $q$ is, because high-$q$ GHE's are more strongly affected by rare and large events. The average error for $H^{250}_{1}$ for the entire time-line is $0.028\pm0.014$ and the respective error for $H^{250}_{4}$ is $0.034\pm0.017$. As it is evident from figures~\ref{fig:SP500_RAW_250}~and~\ref{fig:SP500_RAW_750}, the two {\it w}GHE's of the randomly generated surrogates are evenly distributed around 0.5, the expected value for a random series, whereas the the two {\it w}GHE's of the real S\&P~500 data clearly depart from these values. We also notice that, for the surrogate series, $H_1^{(\theta)}$ and $H_4^{(\theta)}$ evolve almost parallel to each other and are close to each other at all times, as expected for a uniscaling series. However, the two {\it w}GHE's of the real time-series clearly differ at certain time periods and, at some periods, they even follow completely different trends. Notice that there are certain points in the series where $H^{(\theta)}_4$ shows an abrupt drop relative to $H^{(\theta)}_{0.1}$, a drop which decays with a characteristic time that is proportional to $\theta$. These correspond (as we will discuss later in this paper) to large tail events (big rises or drops) that bias the value of the high $q$ \textit{w}GHE. This biasing effect carries on in the future for a characteristic time proportional to $\Delta t$, the length of the averaging window for the \textit{w}GHE calculations and is also dependent on $\theta$. This fact demonstrates why it is highly desirable to choose a $\Delta t$ value as small as possible so that we avoid masking the true multiscaling for a prolonged time in the future of such large tail events, as long as finite size effects are also kept at an acceptably low level.

Table \ref{tab:MeanH} shows the mean values of $H_{1}$ and the extreme {\it w}GHE's, $H_{0.1}$ and $H_4$, as well as $\mathbb{E}\left[|W_{0.1,4}|\right]$, the mean of the absolute value of the difference between the extreme-$q$ {\it w}GHE's of the S\&P~500 series and its respective surrogate. The mean values are calculated over the entire history of S\&P~500 ($\approx1929$ until Feb.~14, 2020) using Equations (\ref{eq:mean}) and (\ref{eq:W}). The standard error of the mean which depends on the uncertainty of determining each value of the exponent $H_q(t)$ from the GHE method is also shown preceded by $\pm$ sign. The standard deviations of the S\&P~500 and its respective surrogate time-series, calculated by Equation (\ref{eq:stdev}), are also reported on a separate line together with their standard errors. The higher the value of $\mathbb{E}\left[|W_{0.1,4}|\right]$ or $B$, the more multiscaling the financial time-series is (on the overall). We see that the mean values of {\it w}GHE for $q=1$ of the real S\&P~500 data are higher than 0.5 within standard error. The mean values of {\it w}GHE for $q=4$ are lower than 0.5 within standard error. Also, the mean absolute value of $W_{0.1,4}$ and $B$ are also greater than 0, within standard error. All the above imply that, on the average, the S\&P~500 index, during in its entire historical time span is characterised by multiscaling. Also, the fact that $H_1$ is also statistically greater than 0.5, suggests that S\&P~500 has been, historically and on the average, a slightly persistent market. On the other hand, the average values of the respective quantities for the randomly generated surrogate S\&P~500 time-series show that, on the average, all the Hurst exponents of a random time series with a varying volatility profile that matches that of the real series, shows neutral behavior. The above results agree with previous studies of the Hurst exponent of the S\&P~500. 

\begin{table}[ht!]
\centering
\caption{\label{tab:MeanH} Statistics of {\it w}GHE's for the SP500 index: comparison between real and surrogate data. $\theta=750$ days.}
\vspace{2pt}

\begin{tabular}{|c|c|c|c|c|c|}
\hline
                     & $H_{0.1}^{(\theta)}$ & $H_1^{(\theta)}$ & $H_4^{(\theta)}$ & $\mathbb{E}\left[|W_{0.1,4}|\right]$ & $B$ \\ \hline
\textbf{S\&P~500}      &        \makecell{$0.5511$  \\ $\pm 4.7x10^{-4}$}      & \makecell{$0.5244$  \\ $\pm 3.2x10^{-4}$}        &     \makecell{$ 0.4454$ \\ $\pm 4.9x10^{-4}$}         &   \makecell{ $0.1058$ \\ $\pm 6.8x10^{-4}$}                           & \makecell{  $-0.02951$ \\ $\pm 8.2x10^{-6}$} \\ \hline
\textbf{(Standard deviation)} &  \makecell{$0.06101$ \\ $\pm 7.1x10^{-6}$}       &    \makecell{ $0.05671$ \\ $\pm 4.8x10^{-6}$}       & \makecell{$0.06623$ \\ $\pm 8.8x10^{-6}$}     & \makecell{ $0.09003$ \\ $\pm 1.1x10^{-5}$} &  \makecell{ $0.020028$ \\ $\pm 2.0x10^{-7}$}  \\ \hline
\textbf{Surrogate data}       &      \makecell{$0.4943$ \\ $\pm 3.7x10^{-4}$}       & \makecell{$0.4940$ \\ $\pm 2.5x10^{-4}$}       &  \makecell{    $0.4853$ \\ $\pm 3.1x10^{-4} $}       & \makecell{  $0.0089$ \\ $\pm 4.8x10^{-4}$}                            &   \makecell{$-0.001976$ \\ $\pm 4.8x10^{-7}$}  \\ \hline
\textbf{(Standard deviation)} &   \makecell{   $0.03615$ \\ $\pm 5.7x10^{-6} $ }  & \makecell{$ 0.03472$ \\ $\pm 2.0x10^{-5}$}   &  \makecell{     $0.04324$ \\ $\pm 5.0x10^{-6}$}  &  \makecell{  $0.05636$ \\ $\pm 7.6x10^{-6}$}  &   \makecell{ $0.016943$ \\ $\pm 3.2x10^{-7}$}  \\ \hline
\end{tabular}
\end{table}

Turning to the temporal evolution of $H_q$ for various values of $q$, it is already apparent from figures~ \ref{fig:SP500_RAW_250}~and~\ref{fig:SP500_RAW_750} that persistence and multiscaling may vary with time as there are time periods when the index seems to be persistent, others when it is neutral and others when it is anti-persistent. Similarly, there are time periods when it is multiscaling and others where it is uniscaling indicated by the relative deviation between the $H_{0.1}$ and $H_4$. There are also some time periods where $H_{0.1}$ and $H_4$ seem to evolve with similar local trend and some time periods where they seem to follow different or even opposite trends. The later signifies an anomalous kind of $H_q$ profile evolution that is probably related to particular changes in the underlying dynamics of the market. In order to investigate these matters in more detail, we perform the analysis described in the next paragraphs.

First, we apply a 2nd-order polynomial smoothing filter to $H_q^{'(\theta)}$ data for a time window of length equal to 240 trading days (~ 1 year) in order to reduce the noise and more clearly identify the underlying temporal patterns in the GHE spectra. 

\begin{figure}[h!]
\includegraphics[scale=0.35]{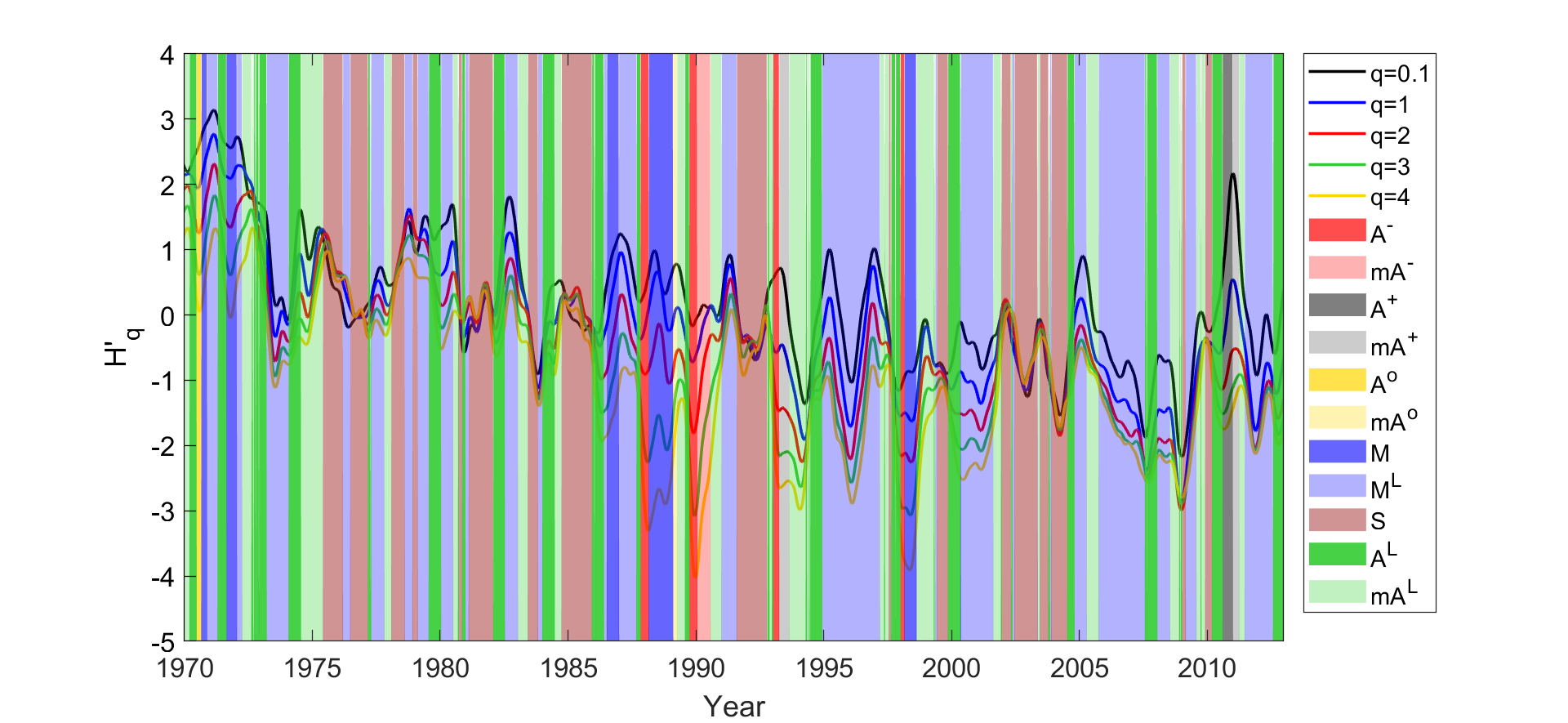}
\caption{\label{fig:PatternDefinition} Temporal patterns in \textit{w}GHE time-series for the S\&P~500 index for the period Jan. 2, 1970 to Jan. 2, 2013.}
\end{figure}

Next we inspect the smoothed $H_q^{'(\theta)}$ series for S\&P~500 log prices, and the five values $q=0.1,1,2,3,4$, as shown in figure~\ref{fig:PatternDefinition}. We identify several distinct \textit{Temporal Patterns} (TP) in the co-evolution of the series of the extreme $q$'s ($q=0.1, 4$) based on: 
\begin{enumerate}
    
    \item The standardized 'width' $W'_{0.1,4}$ of the {\it w}GHE $q$-spectrum, as defined in Equation (\ref{eq:Wnorm}), which, as already said, is a measure of the multiscaling of the index at time $t$. Looking at the co-evolution of the five $H'_q$'s shown in figure~\ref{fig:PatternDefinition}, we distinguish time periods where the five $H'_q$'s are very close to each other (signifying strongly uniscaling behavior), time periods where the five $H'_q$'s are clearly apart (signifying time periods of multiscaling) and time periods where they are strongly diverging, i.e. time periods where multiscaling is stronger. Therefore, we define three different levels of multiscaling by comparing $W'_{0.1,4}$ to two threshold values $\phi$: $\phi_L$ and $\phi_H$ which correspond to the low and high threshold values. If $W'_{0.1,4}(t)>\phi_H$ we consider that the index is characterized by strong multiscaling (denoted either by letter M or A), while, for small widths $W'_{0.1,4}(t)\lesssim\phi_L$), it is characterized as uniscaling (denoted by letter 'S'). For intermediate widths $\phi_L\lesssim W'_{0.1,4}(t)\lesssim\phi_H$ we characterise it as 'weak multiscaling' and denote it with the letter M$^L$ (or A$^L$). Similarly, in the case we measure multiscaling by using the $B$-proxy instead of $W$, we use the standardized value $B'$ as defined in Equation (\ref{eq:Bnorm}) and compare it to $\phi_L$ and $\phi_H$. $B'>\phi_H$ denotes a strong multiscaling pattern (M or A) and $B'<\phi_L$ a uniscaling S pattern, whereas $\phi_L\lesssim W'_{0.1,4}(t)\lesssim\phi_H$ a 'weak multiscaling' M$^L$ or A$^L$ pattern.
    \item The difference between the 'local trends' of the extreme \textit{w}GHE curves $H'_{0.1}(t)$ and $H'_4(t)$ at time $t$. The local trends could be defined as the time derivative of the \textit{w}GHE series at time $t$, but in order to get a statistically significant measure we use the Change Point Analysis method (CPA), as will described later in the paper. We denote by letter M (stands for 'muliscaling') or M$^L$ (stands for 'low' multiscaling), a wide TP, as determined by the procedure described in the previous point, for which the local trends are statistically equal. In an M (or M$^L$) pattern, the extreme $H_q$ time-series move parallel to each other and thus the width $W$ remains statistically unchanged. Conversely, we denote by letter A (or A$^L$) (stands for 'asymmetric' multiscaling) a wide TP, in which the extreme {\it w}GHE's evolve in statistically different directions and/or different rates.
    \item The 'asymmetry' in local trends: 'A' patterns can come in the following three variations: (i) an A$^-$: a TP in which $H_4$ drops at a rate faster than $H_{0.1}$ either drops or rises; (ii) A$^+$: a TP in which $H_{0.1}$ rises at a rate faster than $H_4$ either drops or rises; (iii) A$^0$ a TP in which $H_{0.1}$ rises at approximately the same rate as $H_4$ drops; (iv) mA$^-$ is a pattern in which $H_4$ rises at a rate faster than $H_{0.1}$ either drops or rises\footnote{The prefix $m$ stands for mirror image of the pattern.}; (v) mA$^+$ is a pattern in which $H_{0.1}$ drops at a rate faster than $H_4$ either drops or rises; (vi) mA$^0$ is a pattern in which $H_{0.1}$ drops at approximately the same rate as $H_4$ rises. For the 'weakly multiscaling' asymmetric TP's A$^L$, we do not define any '+' or '-' TP's, just the diverging TP A$^L$ and the converging (mirror) TP $mA^L$.
    \item The relative variation among the GHE's across the $q$ values, \textit{e.g.} the ordering of the GHE's \textit{vs.} $q$ at a particular time instance. Specifically, in some time periods the concavity relation can be violated giving way to a \textit{'reversed' TP}, in which {\it w}GHE's of higher $q$'s are larger than {\it w}GHE's of smaller $q$'s. We denote such TPs by attaching the prefix 'r' to the symbols of any of the above TPs. It is important to highlight that 'reversal' is a particularly rare phenomenon as it entails the effect for which dependence would be stronger for tail events than for common events. 'Reversal', however, is realistically expected for severe crisis periods, where the price change distribution strongly deviates from a Gaussian distribution and tail events are very frequent and highly correlated.
    \item  The transition state from a uniscaling period to a multiscaling period and vice-versa. In case we have a weakly multiscaling TP for which the extreme wGHE's seem to diverge tending to turn into a multiscaling pattern, then, if at a particular time $t$, $W'_{0.1,4}>\phi_L$ (or $B'>\phi_L$) and the local trends of the extreme \textit{w}GHE's are statistically \textit{diverging}, we define a 'transition', weakly-multiscaling TP, that we denote by A$^L$. If, on the other hand, the local trends of the extreme \textit{w}GHE's are statistically \textit{converging}, then define a 'transition' weakly-multiscaling TP that we denote by $mA^L$, i.e. the 'mirror' A$^L$.
\end{enumerate}

\begin{figure}[h!]
\includegraphics[scale=0.17]{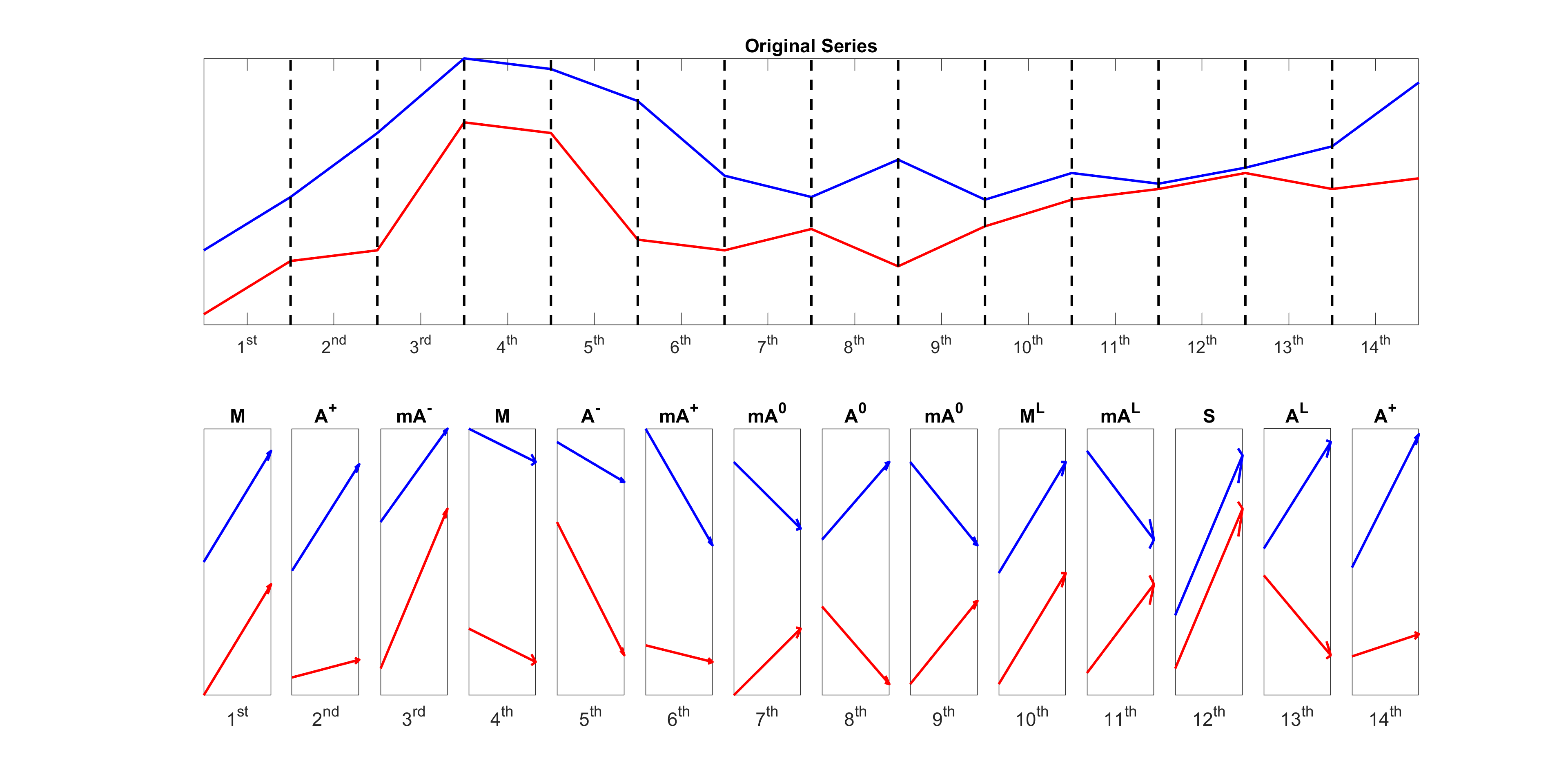}
\caption{\label{fig:PatternRecognition} Schematic depiction of GHE TP's. In the upper plot, a schematic representation of two $H_q$ time-series is shown for the extreme $q's$. Blue color represents the minimum $q$ series and the red color the maximum $q$ series. In the lower plot, the respective TP's are labeled.}
\end{figure}

In figure~\ref{fig:PatternRecognition} we summarize and schematically present the TP's described above.

\subsection{TP identification algorithmic procedure}
In this section we present the algorithmic procedure to extract the TPs from {\it w}GHE's series in a statistically rigorous way. The procedure contains the following steps:

\begin{enumerate}
    \item First, select the standardized metric $\gamma'$ to be used as a measure of multiscaling: $\gamma'=W'_{q_1,q_2}$ or $\gamma'=B'_{q'_1,q'_2}$, as well as the respective pair of extreme $H'_{q_1}$ and $H'_{q_2}$ time-series that will be used for determining the local trends. Select a $\theta$ value and a sliding window length $\Delta t$ and compute the relative \textit{w}GHE's for both the real series and a random surrogate, with log-returns drawn from a normal distribution standard deviation equal to the volatility profile of the real series. Compute the relevant standard deviations of the surrogate series from Equations~(\ref{eq:stdev}) and (\ref{eq:Stdpooled}) and obtain the standardized series from Equations~(\ref{eq:NormH}), (\ref{eq:Wnorm}) or (\ref{eq:Bnorm}). If $\gamma'$ is set equal to $B'_{q'_1,q'_2}$, then compute several series $H_{q}$ between the chosen extreme values. In this work we calculated a set of $H_{q}$'s in the range $q=0.1-1$ with extreme series the ones for $q'_1=0.1$ and $q'_2=1$. Then, for each each time $t$, apply a linear least squares fit to the data $H_q(t)$ \textit{vs.} $q$, the slope of which is equal to $B(t)$. For $\gamma'= W'_{q_1,q_4}$, we chose $q_1=0.1$ and $q_4=4$, in the present work.
    
    \item Smooth out the computed raw standardized series using a 2nd order polynomial smoothing function. We used a smoothing window of 48 data points which (for a skipping window of 5 trading days that we used for \textit{w}GHE calculations) corresponds to 240 trading days, \textit{i.e.} approximately one calendar year.
    
    \item Apply the Change Point Analysis algorithm (CPA) (\cite{killick2012optimal}) to the two extreme series $H'_{q_1}$ and $H'_{q_2}$, in order to get time intervals characterized by the same local trend (rate of increase of the {\it w}GHE's) as well as to obtain the values of the trends. The same or different segment limits can be chosen (same 'binning') for the two series. If the same binning is selected (we chose this option in the present work), then, in practice, CPA is applied to one of the two series (or alternatively to the $\gamma$ series), and the automatically extracted bin limits are then enforced on the application of CPA to the other series. The CPA analysis breaks the series into a set of several segments of potentially different lengths $\{\Delta t_i\}$, and outputs a unique slope value, $\beta_i^{q_1}$ and $\beta_i^{q_2}$ for each segment $i$ and for the respective standardized wGHE series for $q_1$ and $q_2$. 
    
    \item For each data point at time $t$, statistically determine the degree of multiscaling by checking the statistical significance of $\gamma'$ against the predefined threshold value $\phi_L$ and $\phi_H$: if $\gamma'>\phi_H$, then the dynamics of the underlying series is multiscaling (M-type or the various A-type TP's), else if $\gamma'<\phi_L$, it is characterised as uniscaling (S), else it is characterised as 'weakly multiscaling' (M$^L$-type or A$^L$/$mA^L$-type TP's). $\phi_L$ should, in general be much smaller than 1 and $\phi_H$ greater than 1. In the present work, we use $\phi_L=0.32$ and $\phi_H=1.64$ as threshold values, which correspond to the $25^{th}$ and $95^{th}$ percentile of the Gaussian distribution in respect. Other choices are of course possible. The rationale behind the particular choices is that the limit for true uniscaling should be considerably \textit{lower} than the 'noise level' of $W$, as defined by 1 standard deviation of the random surrogate, while the limit for strong multiscaling should be significantly \textit{higher} than the noise level. Therefore, $\phi_L=0.32$ signifies that only 25\% of the widths $W'$ in a random time-series are below this threshold and thus the scaling of the real series at any point in time for which $W'$ is smaller than this value can be characterized as uniscaling in a statistically significant manner. Similarly, $\phi_H=1.64$ signifies that only 5\% of the widths in the random time-series are above this limit, therefore the scaling of the real series at any point in time for which $W'$ is greater than this value can be characterized as multiscaling in a statistically significant manner. Finally, for times when $W'$ values are between these values the scaling is characterized as "weak' multiscaling.
    
    \item For each data point on day $t$, compare the relative slopes of the extreme {\it w}GHE series, as extracted by CPA, at the time bin $i$ to which $t$ belongs, in order to detect the different forms of multiscaling, i.e. to identify whether the TP is an M-type or an A-type. In particular, to designate an A pattern, we require that the absolute difference in the slopes should be $\phi_S$ standard deviations above 0, i.e.:
    
    \begin{equation}
    \label{eq:AvsMcheck}
        \frac{|\beta_i^{q_1}-\beta_i^{q_2}|}{\sigma(|\beta_{surr}^{q_1}-\beta_{surr}^{q_2}|)}>\phi_S,
    \end{equation}
    
    where $\beta_i^{q_1}$, $\beta_i^{q_2}$ are the slopes of $H_{q_1}$ and $H_{q_2}$ at bin $i$, and $\beta_{surr}^{q_1}$, $\beta_{surr}^{q_2}$ is the respective pair of slopes computed on the surrogate data, $\sigma(...)$ denotes the standard deviation of the series and $\phi_S$ is the threshold of the evaluation.\footnote{In general, one could use different thresholds $\phi$ between the width test involved in multiscaling \textit{vs.} uniscaling characterizations, and the slope tests involved in the type of multiscaling characterizations. In the present work, we chose the same value $\phi_S=1.64$ (as $\phi_H$) for all tests.} In other words, this formulation returns an 'A' pattern only if it is statistically greater than the variability of the local trends of the surrogate index, as measured by applying a similar CPA procedure to the extreme series, $H_{q_1}$ and $H_{q_2}$ of the random surrogate data. In practice, one can use the same binning on the surrogate data as determined by the CPA of the real data, which is what we did in the present work.
    
    \item If $W'_{q_1,q_2}>\phi_H$, then distinguish among the various $A$-type multiscaling TP's. At first, compare the absolute value of the difference of absolute values of the local trends $\beta_i^{q_1}$ and $\beta_i^{q_1}$. We check its statistical significance by
    
    \begin{equation}
    \label{eq:AvsAminuscheck}
        \abs*{ \frac{|\beta_t^{q_1}|-|\beta_t^{q_2}|}{\sigma(|\beta_{surr}^{q_1}|-|\beta_{surr}^{q_2}|)}}>\phi_S.
    \end{equation}
    
 If this condition is false, then the TP is an A$^0$. If it is true, then it is either an A$^-$ or A$^+$ or one of their respective mirrors mA$^-$, mA$^+$. In the later case, in order to determine which one of the four, compare the absolute values of the two $\beta$'s and also use the sign of each $\beta$. Specifically: 
 \begin{itemize}
     \item[i] if $|\beta_t^{q_1}|<|\beta_t^{q_2}|$ and $\beta_t^{q_2}<0$, it is A$^-$,
     \item [ii] if $|\beta_t^{q_1}|<|\beta_t^{q_2}|$ and $\beta_t^{q_2}>0$, it is mA$^-$,
     \item [iii] if  $|\beta_t^{q_1}|>|\beta_t^{q_2}|$ and $\beta_t^{q_2}>0$, it is A$^+$
     \item [iv] $|\beta_t^{q_1}|>|\beta_t^{q_2}|$ and $\beta_t^{q_2}<0$, it is mA$^+$.
 \end{itemize}  
    \item If $\phi_L<W'_{q_1,q_2}<\phi_H$, then distinguish between the A$^L$ TP and the $mA^L$ TP, the first corresponding to diverging weakly multiscaling asymmetric patterns and the second to a converging one. The first often precedes a transition between an uniscaling state (S) to an M or M$^L$ multiscaling state. The second precedes the reverse transition, \textit{i.e.} from a multiscaling to a uniscaling state. The condition is that the relative trends $\beta_t^{q_1}$ and $\beta_t^{q_2}$ are sufficiently different, i.e. they satisfy condition~(\ref{eq:AvsMcheck}) and:
     \begin{itemize}
        \item[i] $\beta_t^{q_1} > \beta_t^{q_2}$, then the TP is an A$^L$ else
        \item[ii] $\beta_t^{q_1} > \beta_t^{q_2}$, then it is an $mA^L$.
     \end{itemize}
    \item In case of 'reversal', i.e. if $H'_{q_1}<H'_{q_2}$: then one must simply interchange $H'_{0.1}$ with $H'_{4}$ in the equations presented in all the above points. The resulting TP's will be the 'reverse' TP denoted by an extra letter 'r' in front of the respective TP symbol.
\end{enumerate}

\begin{figure}[h!]
\begin{center}
\includegraphics[height=0.6\textheight,width=1.05\textwidth]{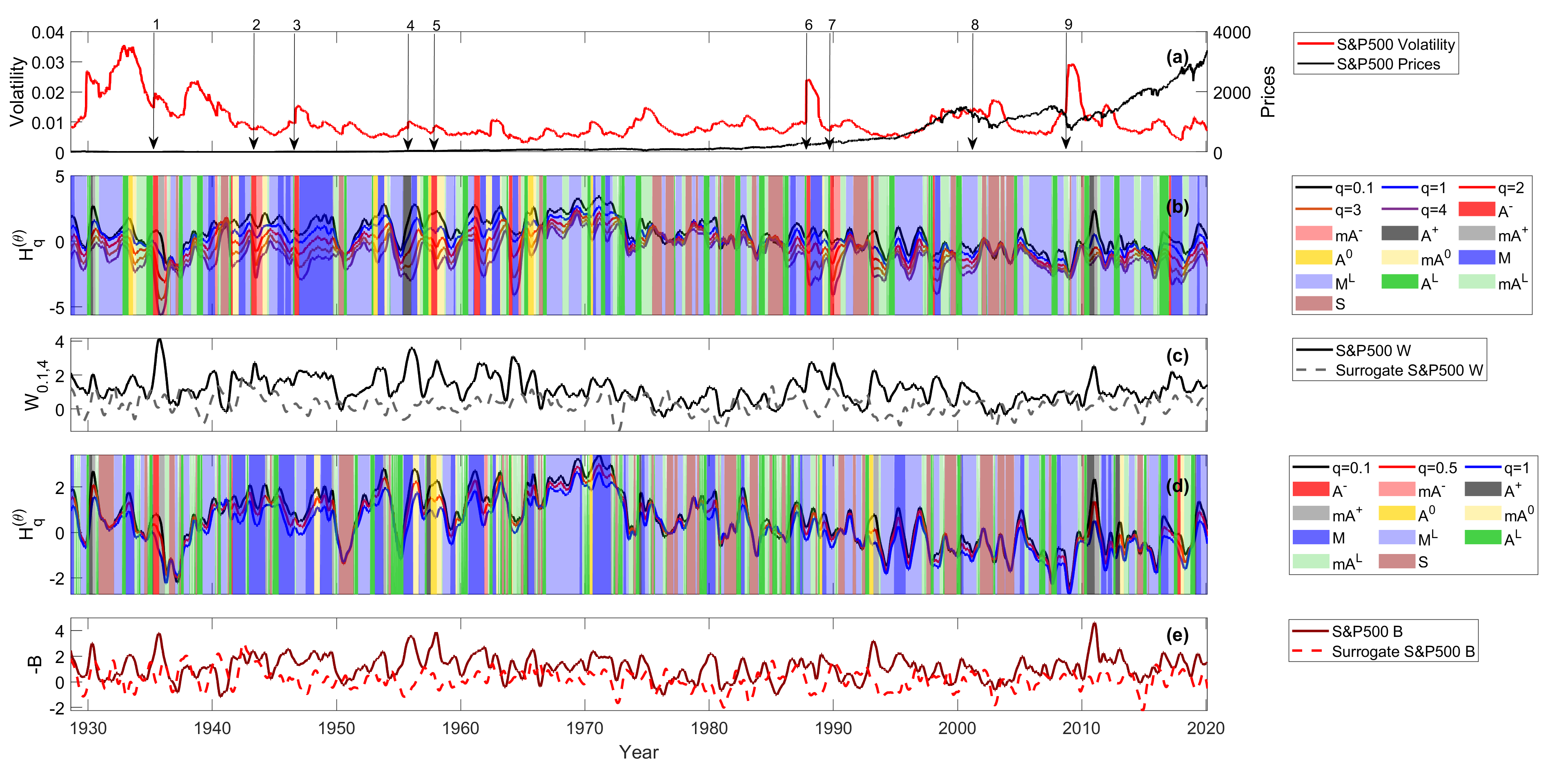}
\caption{\label{fig:SP500close} SP500 index price time-series and scaling TPs: (a) Index closing prices and weighted volatility (b) Normalized \textit{w}GHE's for $q=0.1,1,2,3,4$ with identified TPs using $H(0.1)$, $H(4)$ and $\gamma=W_0.1,4$. (c) Width $W'_{0.1,4}$ of the S\&P~500 normalized GHE's for the real index data and the respective surrogate data. (d) Normalized \textit{w}GHE's for $q=0.1,0.5,1$ with identified TPs using $H(0.1)$, $H(1)$ and $\gamma=-B$. (e) $B$ proxy of the S\&P~500 normalized GHE's for the real index data and the respective surrogate data. Numbered arrows show particular market events, as referenced in text.}
\end{center}
\end{figure}

\begin{figure}[h!]
\begin{center}
\includegraphics[height=0.5\textheight,width=1.05\textwidth]{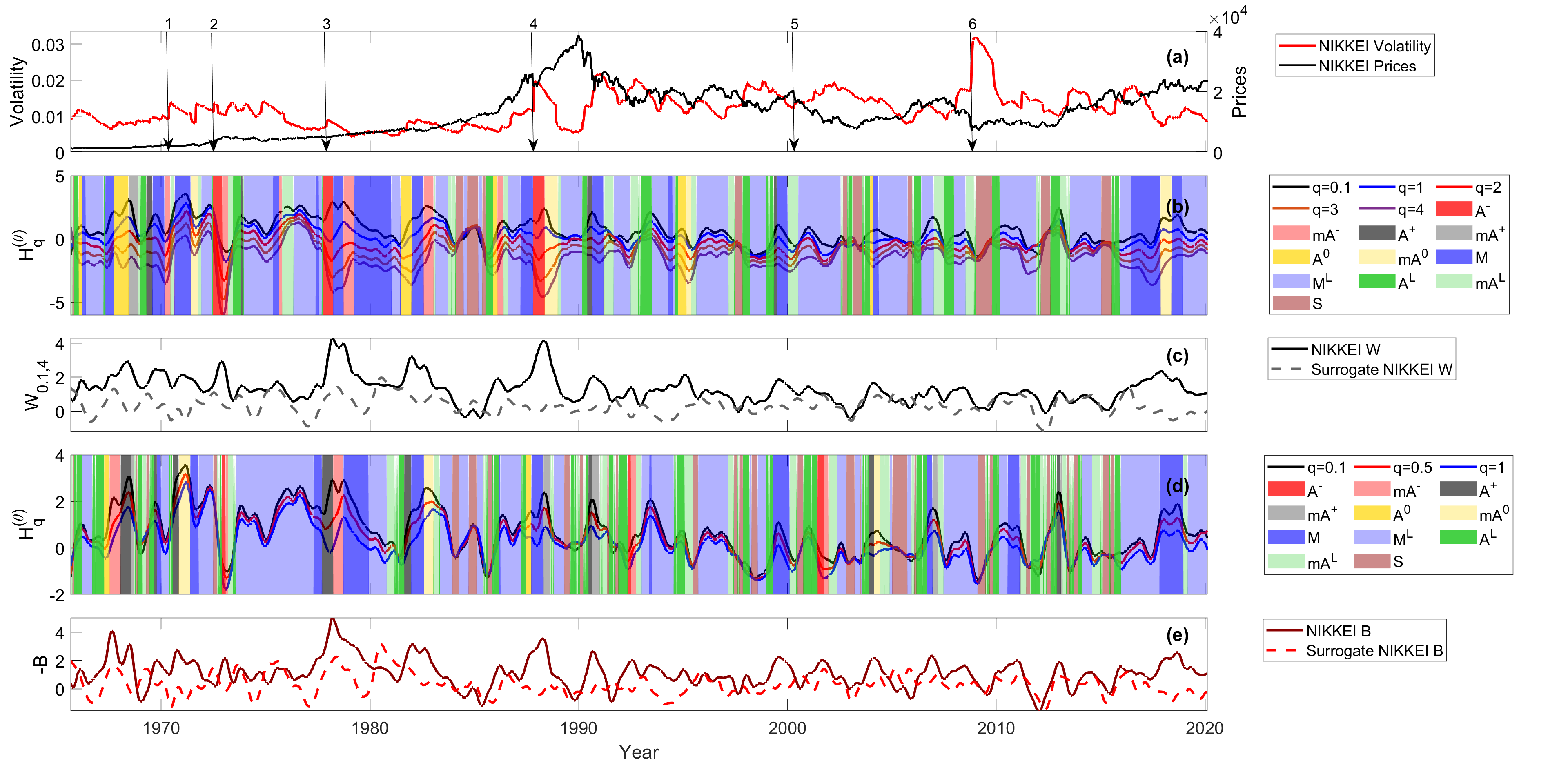}
\caption{\label{fig:NIKKEIclose} NIKKEI index price time-series and scaling TPs: (a),(b),(c),(d) and (e) exactly as described in caption of figure~\ref{fig:SP500close}. Numbered arrows show particular market events, as referenced in text.}
\end{center}
\end{figure}

\begin{figure}[h!]
\begin{center}
\includegraphics[height=0.6\textheight,width=1.05\textwidth]{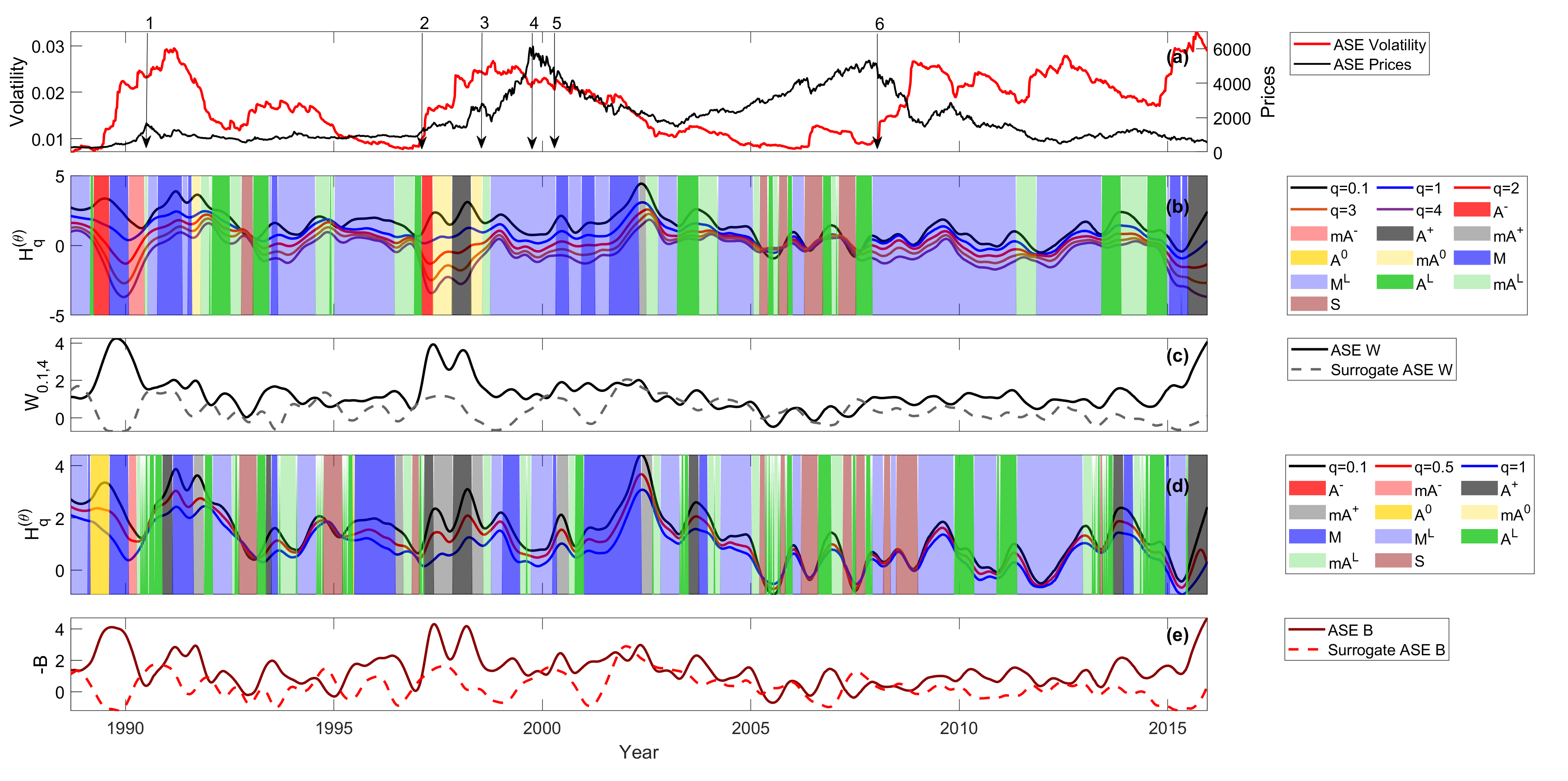}
\caption{\label{fig:ASEclose} ASE  index price time-series and scaling TPs: (a),(b),(c),(d) and (e) exactly as described in caption of figure~\ref{fig:SP500close}. Numbered arrows show particular market events, as referenced in text.}
\end{center}
\end{figure}

\begin{figure}[h!]
\begin{center}
\includegraphics[height=0.6\textheight,width=1.05\textwidth]{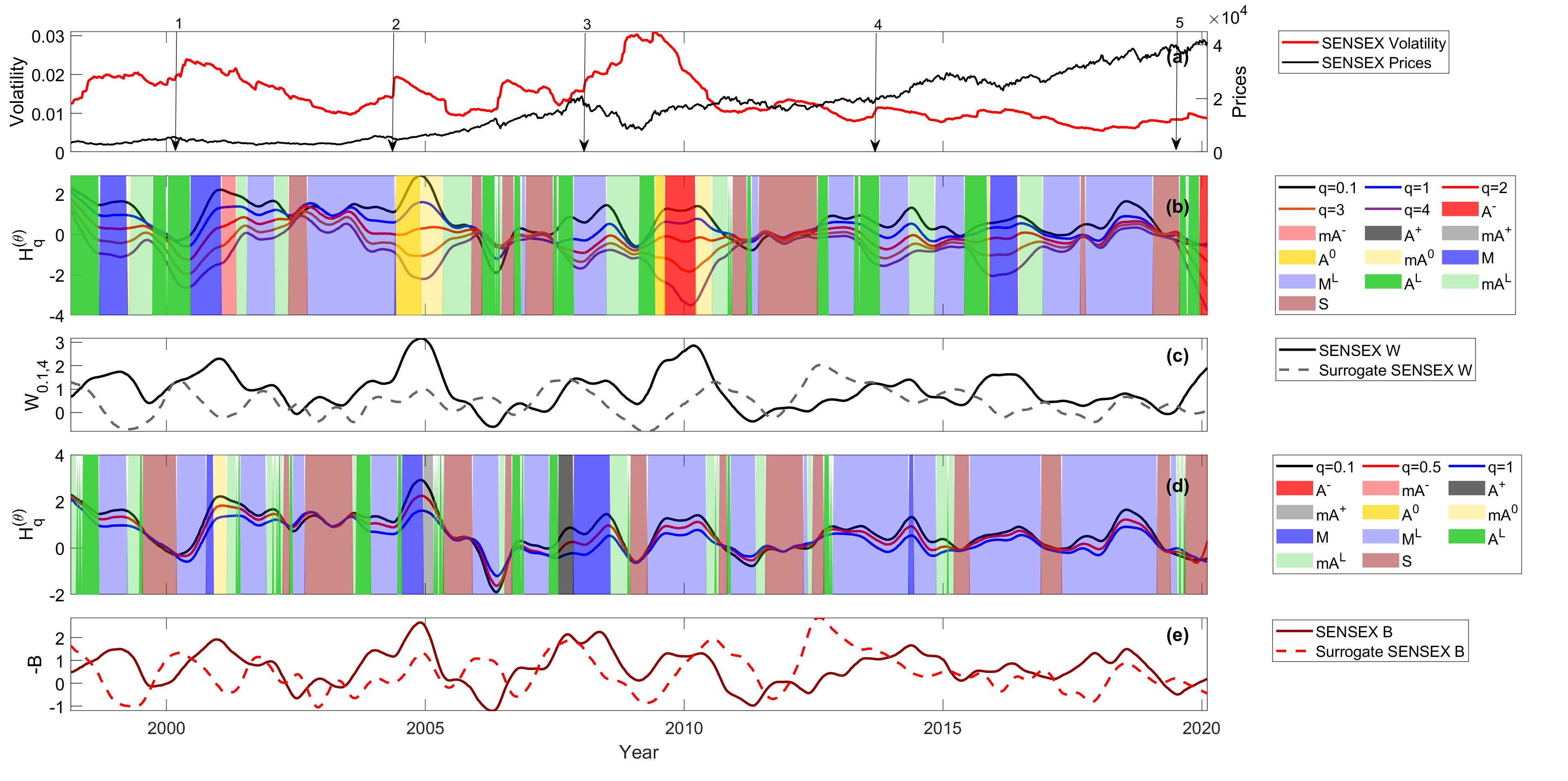}
\caption{\label{fig:SENSEXclose} SENSEX index price time-series and scaling TPs: (a),(b),(c),(d) and (e) exactly as described in caption of figure~\ref{fig:SP500close}. Numbered arrows show particular market events, as referenced in text.}
\end{center}
\end{figure}

The results of the TP identification analysis presented above, as applied to the S\&P~500, NIKKEI, ASE and SENSEX indices are shown in figures~\ref{fig:SP500close}-\ref{fig:SENSEXclose}. In each of these figures the following are plotted: In (a) the weighted volatility series of the index (left axis), calculated by Equation (\ref{eq:volatility}), and the index close prices (right axis); in (b) the normalized \textit{w}GHE's $H'(q)$ time-series of the index for $q=0.1,1.2.3.4$, where we have marked the identified TP's by setting $\gamma'=W'_{0.1,4}$ and using $H'_{0.1}$ and $H'_{4}$ for the extreme \textit{w}GHE's. TP's are marked by color mapping; in (c) the time evolution of the normalized \textit{w}GHE width $W'_{0.1,4}$ together with the width of the respective surrogate index; in (d) the $H'(q)$ time-series for $q=0.1, 0.5, 1$ with the identified TPs by setting $\gamma=-B$ and using $H'_{0.1}$ and $H'_{1}$ for the extreme \textit{w}GHE's. TPs are also marked by the same color mapping as in plots (b). Finally, in (e) we show the B proxy time-series of the real index together with the respective B proxy of the surrogate index.

By examining figures~\ref{fig:SP500close}~-~\ref{fig:SENSEXclose}, we notice several interesting facts:

\begin{enumerate}
    \item As clearly seen from plots (b) and (c), the scaling of the various indexes varies significantly with time: there are certain time periods when $W'_{0.1,4}$ is much higher than the average width of the GHE spectrum of the surrogate index, signifying a definite multiscaling structure of the underlying dynamics, while there are time periods when $W'_{0.1,4}$ is very small, signifying a uniscaling structure. Moreover, the transition between a period of multiscaling behavior to a period of uniscaling behavior can be rather sharp, a fact that alludes to the existence of transition occurring in the underlying index dynamics.
    \item There are time periods of persistent behaviour ($H_1^{'(\theta)}>0$), time periods of anti-persistent behavior ($H_1^{'(\theta)}<0$) and time periods of neutral behavior ($H_1^{'(\theta)} \approx 0$). If one generalizes the notion of 'persistence' to include GHE's of $q$ values different from 1, then there are time periods when the small $q$ GHE's rise or stay approximately the same, while $H'_4$ is dropping, \textit{i.e.} moving in the opposite direction to a more 'anti-persistent' scaling. This behavior, which is characterised by A$^0$ or A$^-$ TPs, is connected to one or more isolated, large price change events (tail events) that occur in a direction opposite to the local market trend (\textit{e.g.} a large price drop in an otherwise rising market or vice-versa). A notable example is the 'Black Monday' event that occurred on Monday, Oct. 19th 1987 (and Tuesday Oct. 20 in some markets), where S\&P~500 (arrow~6), for example, lost more than 20\% in one day. The event was followed by a large rise in the next day and the index made up for all the losses soon after. 'Black Monday' occurred amidst a bullish market period and was similarly followed by a rising trend. A single tail event of this size causes a large bias, especially in the high $q$ \textit{w}GHE's, hence the pronounced A$^-$ TP is observed, as appears in figures~\ref{fig:SP500close}b and \ref{fig:NIKKEIclose}b for both S\&P~500 and NIKKEI (arrow~No~4). Such TP's are also seen after the 'Asian' and the 'Russian' related crises drops in 1997 and 1998\footnote{For the related dates of these isolated market drops, see table~\ref{tab:DeletedDates}.} in respect, which also occurred amidst a rising market and in several other occasions along the index price timeline such as, for example: (i) S\&P~500: April, 16-17, 1935 when a $\approx 9\%$ drop is followed by a $\approx 9\%$ rise, May, 16-17, 1935 when a $\approx 7\%$ drop was followed by a $\approx 9.4\%$ rise and August 16,19, 1935, when a $\approx 8\%$ drop was followed by a $\approx 7\%$ rise (arrow~No~1 in figure~\ref{fig:SP500close}). (ii) NIKKEI: June, 26-27, 1972 when a $\approx 8\%$ drop was followed by a $\approx 5.3\%$ rise (arrow~No~2 in figure~\ref{fig:NIKKEIclose}), Jun., 26-27, 1972 when a $\approx +4\%$ rise occurs amidst a dropping trend (arrow~No~3 in figure~\ref{fig:NIKKEIclose}). Notably, these A$^-$ patterns are not present in the B-proxy series shown in plots (d) of the said figures, since the small $q$ \textit{w}GHE's are not so much affected by tail events, except for the 1935 large A$^-$ TP for S\&P~500 which appears there too, because that particular TP was caused by several more than one big tail events over an extended period of time. However, there are time periods when the small $q$ \textit{w}GHE's show a sharp rise while the $H'_4$ drops or is almost unchanged (a behavior that yields an A$^+$ TP). This behavior hints to a situation where a one or more large events occur in the same direction as the current market trend, meaning that they give a large 'persistent' boost in the small $q$ $H'_q$'s. An example of the latter behavior is in the 2008 real estate crisis (arrow~No~9 in figure~\ref{fig:SP500close}), an exogenous to stock-market event, where large index price drops occurred amidst a period of a rapidly falling market, evidence of persistent 'herding' behavior following the 2008 crash. These few large drops cause a sharp rise in $H'_{0.1}$ and $H'_1$, rather than a drop in $H'_4$. The same pattern is seen in 2012 for S\&P~500, where the observed $A+$ TP is a result of big daily rises amidst a rising market period. One more example of an A$^+$ TP coming from several big rising events amidst a period of rising rising market trend is the one shown by arrow~No~4 in figure~~\ref{fig:SP500close}, in period May,~1955-Sep.~1956. Finally, there are other time periods when the scaling is consistently persistent (or anti-persistent, or neutral) for all values of $q$, meaning that either the period is void of large rising or dropping events and/or that both large and small events follow the same scaling behavior.
    \item Multiscaling behavior is not necessarily correlated with periods of increased index volatility or periods of persistent scaling: there are time periods showing both high volatility and multiscaling/persistent behaviour, as well as time periods with low-volatility and multiscaling/anti-persistent behaviour. Time periods when volatility and multiscaling are positively correlated, include those which contain a single extreme market drop tail event, which is sufficiently large to impact both volatility and GHE calculations. An example of this fact is seen in the period 1987-1988, following 'Black Monday'. As an example of a period showing a large increase in multiscaling strength, while volatility remains low, we mention the first semester of 1993 for S\&P~500. During this period, we observe a type of asymmetric multiscaling which is the product of a sequence of smaller tail events, distributed over a longer period of time and also depends on how these events are temporarily correlated. Notice also, that A$^-$ TP's caused by a single tail event (that necessarily leads to a sharp volatility rise as well) and A$^-$ TP's caused by temporal correlations and tail events distributed over an extended period of time, also have different shapes: in the first case, the width of the TP decays (following the characteristic rate that depends on the choice of $\theta$), whereas, in the second case, it does not decay immediately, but remains wide for a longer time while the variation of its width does not depend on the choice of $\theta$.  
    \item At the beginning of a bubble, a strong uniscaling behaviour is observed at which the investor heterogeneity seems to be low. Whilst the market starts to grow, the complexity of the time series appears to increase in both measures of multiscaling through an asymmetric TP (usually A$^-$ or A$^L$) and then comes back to uniscaling or moderate multiscaling after the bubble has exploded. This is apparent in both Dot.com bubble and the US real estate bubble, but also in ASE 2000 bubble, in ASE 1990 crash as well as the Japanese 1991 bubble. It is even apparent before the 'Black Monday' crash, both for NIKKEI and S\&P~500, when we notice a clear transition from uniscaling to strong multiscaling via an A$^L$ TP starting back in 1986. In general, before any critical event we necessarily have a transition from uniscaling to multiscaling ranging from a few months to a couple of years before the beginning of the bubble break or crash. It must be noted that the A$^-$/A$^L$/A$^0$ type TP's that we encounter in these transitions are not the 'after-effect' of single large tail market events, but rather the consequence of a transition to multiscaling behavior due to smaller tail events occurring over a prolonged period. Examples of such TP's are: (i) S\&P~500 1956-1957 (arrow~No~5 in figure~\ref{fig:SP500close}), an A$^L$ TP followed by a an A$^0$ TP which was actually followed by a small crash (micro-bubble) at the last quarter of 1957, (ii) S\&P~500 1961, an A$^-$ TP that was followed by a small crash in 1962, (iii) ASE: the pronounced A$^-$-A$^0$-A$^+$-A$^0$ sequence before the ASE big 2000 bubble, as well as the A$^-$-A$^0$ TP before the 1990 crash. (iv) The A$^L$ TP's just before the 2000 'dot.com' bubble in S\&P~500 (arrow~No~8 in figure~\ref{fig:SP500close}), NIKKEI (arrow~No~5 in figure~\ref{fig:NIKKEIclose}) and SENSEX (arrow~No~1 in figure~\ref{fig:SENSEXclose}). (v) The A$^L$ TP's just before the 2008 US real-estate crisis in S\&P~500 (arrow~No~9 in figure~\ref{fig:SP500close}), NIKKEI (arrow~No~6 in figure~\ref{fig:NIKKEIclose}), ASE (arrow~No~6 in figure~\ref{fig:ASEclose}) and SENSEX (arrow~No~3 in figure~\ref{fig:SENSEXclose}). See also, the plots in the Appendix, showing zoomed versions of some particular time periods for S\&P~500 and NIKKEI.
    \item The multiscaling width $W_{0.1,4}$ and multiscaling depth $B$ convey different information in some cases. For example, during the 2008 great financial crisis, $W_{0.1,4}$ doesn't increase too much while $B$ increases sharply. This is because $W_{0.1,4}$ better captures the heterogeneity in the market, which is, to some extent, lower when all investors go in the same direction (selling orders), while $B$ measures the complexity of such heterogeneity, as the distribution of $H_q$ inside a range of q's matters instead of only its boundary values, as we also mentioned above.
    \item Multiscaling does not necessarily imply bad market conditions. When we have multiscaling of type M, it usually reflects good market conditions, even if there is increased heterogeneity (and complexity) in the market.
    \item Multiscaling time periods detected by the GHE spectrum curvature $B$ is on the overall in line with the ones detected with $W$. However, some differences are observed in specific time periods. In particular, during crisis events, the $B$ is more symmetrically multiscaling while $W$ shows many more asymmetries. This is due to the fact that $W$ is affected by the tails of price change distributions considerably more than $B$. In general, it is useful to look at both measures of multiscaling as they emphasize the opposite ends of the price change distribution (small large changes) and thus are complementary to each other. 
\end{enumerate}
 
We now elaborate on some events on each of the indexes analysed. Regarding S\&P~500, depicted in figure~\ref{fig:SP500close}, we can highlight the following facts:
\begin{enumerate}
     \item The Hurst exponent, $H_1$ has a positive trend up to 1971, when it reverses to a long-term negative trend, moving from a persistent signal to a more random signal. This coincides with the end of the Bretton Wood system.
     \item Before the Black Monday of October 1987, the time series presents a uniscaling pattern followed by a moderate asymmetric pattern which is then followed by strong multiscaling. At the same time, the volatility is quite low, meaning that the increased complexity is not driven by a single event but by the market structure. 
     \item Before the Dot.com bubble burst on the second quarter of 2000, we have for both $W_{q_1,q_2}$ and $B$ a sequence of patterns, i.e. converging moderate multiscaling - uniscaling - diverging moderate multiscaling - strong multiscaling. This is accompanied by relatively low but increasing volatility. This is a signal that the market is going to saturate and a probable drop is expected. It can be attributed to the fact that the increasing multiscaling along with a rising volatility increases the market heterogeneity, which is becoming driven by turbulence in trading patterns. 
\end{enumerate}
 
 In figure~\ref{fig:NIKKEIclose} we show another major index, the NIKKEI. Some particular features of this index are: 
 \begin{enumerate}
     \item An uniscaling behaviour at the beginning of 1986 which evolves to an asymmetric multiscaling behaviour of type A$^0$ and A$^-$ and then evolves to a persistent multiscaling, even after the bubble has exploded in 1991.
     \item After the bubble exploded in 1991, the market follows an anomalous scaling. In fact, the market remains moderately multiscaling. This reflects the heterogeneity generated by the monetary policies adopted by the central bank of Japan.
     \item The series appears persistent from 1970 up to the bubble explosion, when a mix of neutral and anti-persistent behaviour are then more present. This behaviour persists up to 2007.
 \end{enumerate}
 
In figure~\ref{fig:ASEclose} we report the analysis related to the Athens stock market. The plots show:
\begin{enumerate}
\item Before the 2000 bubble, an asymmetric multiscaling period is identified, which is a signature of a turbulent period. This is retrieved both using the W and B metrics which remains quite high for the consequent period. This is a combination of the global turbulence in the 1997 and 1998 and the Dot.com bubble which was going to break.
\item From 2005 to the third quarter of 2008 we have a succession of uniscaling and moderate multiscaling patterns which is then followed by a long period of moderate multiscaling pattern, suggesting that a the inception of the global financial crisis a complex dynamic with stronger heterogeneity is taking place.
\item The $B$ proxy agrees almost perfectly with the multiscaling width. This is mainly because, apart from the Dot.com bubble, the turbulent time periods were generated by complex dynamics which increase the heterogeneity of the process in a symmetric way rather than by extreme tail events. 
\end{enumerate}

Finally, we report in Figure~\ref{fig:SENSEXclose} the plots for the Bombay stock market (SENSEX). We observe the following:

\begin{enumerate}
    \item Between the third quarter of 2003 and first quarter of 2004 we see a short uniscaling behaviour followed by a strong multiscaling behaviour. This corresponds to the election of Sonia Gandhi's communist coalition in May 2004 (arrow~No~2) which generated a market drop of 15.52\% and a consequent market heterogeneity in market conditions. In particular, it is possible to notice that the highest and lowest $H_q^{'(\theta)}$ for the $W_{0.01,4}$ go in opposite directions, resulting in a A$^0$ type of pattern. 
    \item $H_1^{'(\theta)}$ is always higher than 0, which implies a persistent behaviour while $H_4^{'(\theta)}$ is, apart few local exceptions, always negative, implying an antipersistent behaviour.
    \item For this index, the $B$ and the $W$ multiscaling proxies disagree in most of the cases. In fact, it is possible notice that in time periods of high width as 2001-2003 and 2011-2012, we have a relatively low $B$, which, together with the high volatility of the series, makes clear that the high width is due more to tail events than temporal correlations.
\end{enumerate}

 It is important to notice that major indices are affected by global events which also spill-over to the peripheral ones, while the opposite is not always true. In fact, it is possible to notice that 'Black Monday', the Japanese bubble and the Dot.com bubble generated a scaling change also in markets different from the ones in which they originated. In contrast, the main shifting events in peripheral indices, do not affect main ones. Given the results depicted in ~\ref{fig:SP500close}~-~\ref{fig:SENSEXclose}, we conclude that a transition from a uniscaling to a multiscaling pattern (usually through an asymmetric pattern\textbf{ of type A$^-$ (strongly multiscaling and asymmetric) or A$^L$ (more weakly multiscaling and asymmetric)} in combination with a relatively low (but rising) volatility is a warning signal that the market is becoming saturated and a turbulent period can follow with a possible crash.

\subsection{Robustness of TP's as warning signals}
\label{robustness}

Having noted all the above, the indication that temporal evolution of multiscaling strength, in both its symmetric and asymmetric forms, as described by the TP's that were defined above and identified by the algorithmic procedure presented in this work, provides possible signals for future market behavior should be further investigated. In order to take the next step, we must first distinguish between the effect on multiscaling coming from the biasing of \textit{w}GHE values caused by tail events which is observed immediately \textit{after} these events, and the effect on multiscaling coming from either tail events or temporal correlations in price changes that occur \textit{prior} to a market crisis, such as a stock market bubble under development, or before a market crash. The first is an 'after-effect' of single and extreme market events, the second is a signal preceding an actual critical event. In an attempt to address the issue, we deleted one or more single trading days from the index time-series that correspond to specific events and recalculated the $H'_q$ profiles of the modified index. More specifically, we deleted some key trading days in S\&P~500, NIKKEI, ASE and SENSEX that are directly related to one or more of these critical events: (i) 'Black Monday', (ii) the 1997 'Asian' crisis, (iii) the 1998 'Russian' crisis (iv) the I. Ghandi election in India in 2004. The exact trading dates that were erased for each index are shown in Table~\ref{tab:DeletedDates} together with the corresponding \% close price changes. Cells containing a dash denote that these dates were not deleted for the particular index. Notice that for each event, we deleted possibly different days and different number of days per index. This is because each market reacted differently to the particular crisis. In particular we wanted to capture and remove the 'instantaneous' effect of a single market event on the GHE computations and the resulting multiscaling, not its possible short or long-term after-effects on the actual market dynamics. Therefore, we deleted just 1 up to 4 trading days that were directly associated with the single market event, usually a large drop followed by a big rise or other smaller rises/drops. 

\begin{table}[!h]
\centering
\begin{tabular}{|c|c|c|c|c|c|}
\hline
\textbf{Deleted dates} & \textbf{Market event} & \textbf{S\&P~500} & \textbf{NIKKEI} & \textbf{ASE} & \textbf{SENSEX} \\ \hline
19-Oct-1987 & B. M.  & -20.47\%  & -    & -  & -   \\ \hline
20-Oct-1987 & B. M.  & +5.33\%  & -14.90\%    & -  & -   \\ \hline
   21-Oct-1987 & B. M.  & +9.10\%  & +9.30\%    & -  & -   \\ \hline
   22-Oct-1987 & B. M.  & -3.92\%  & -    & -  & -   \\ \hline
   23-Oct-1987 & B. M.  & -  & -4.93\%    & -  & -   \\ \hline
   26-Oct-1987 & B. M.  & -  & -4.30\%    & -  & -   \\ \hline
   27-Oct-1997 & A. C.  & -6.87\%  & -    & -  & -   \\ \hline
   28-Oct-1997 & A. C.  & -  &     & -  & -   \\ \hline
   31-Oct-1997 & A. C.  & -  & -    & -4.02\%  & -   \\ \hline
   04-Nov-1997 & A. C.  & -  & -    & +4.72\%  & -   \\ \hline
   06-Nov-1997 & A. C.  & -  & -    & -4.23\%  & -   \\ \hline
   28-Aug-1998 & R. C.  & -  & -3.46\%    & -  & -   \\ \hline
   31-Aug-1998 & R. C.  & -6.80\%  & -    & -  & -   \\ \hline
   01-Sep-1998 & R. C.  & -  & -    & -3.81\%  & -   \\ \hline
   02-Sep-1998 & R. C.  & -  & -    & +5.15\%  & -   \\ \hline
   14-May-2004 & G. E.  & -  & -    & -  & -0.0610 \%   \\ \hline
   17-May-2004 & G. E.  & -  & -    & -  & -11.14\%   \\ \hline
   18-May-2004 & G. E.  & -  & -    & -  & +8.25\%   \\ \hline
\end{tabular}
\caption{Deleted dates for modified indices, the corresponding market event and the \% close price change of that date per index. Cells with a '-' correspond to dates that were not deleted for the particular index in the respective column. 'B.M.' stands for 'Black Monday', 'A.C.' for 'Asian Crisis', 'R.C.' for 'Russian Crisis' and 'G.E.' for Gandhi 2004 Election'.}
\label{tab:DeletedDates}
\end{table}

\begin{figure}[h!]
\begin{center}
\includegraphics[height=0.6\textheight,width=1.05\textwidth]{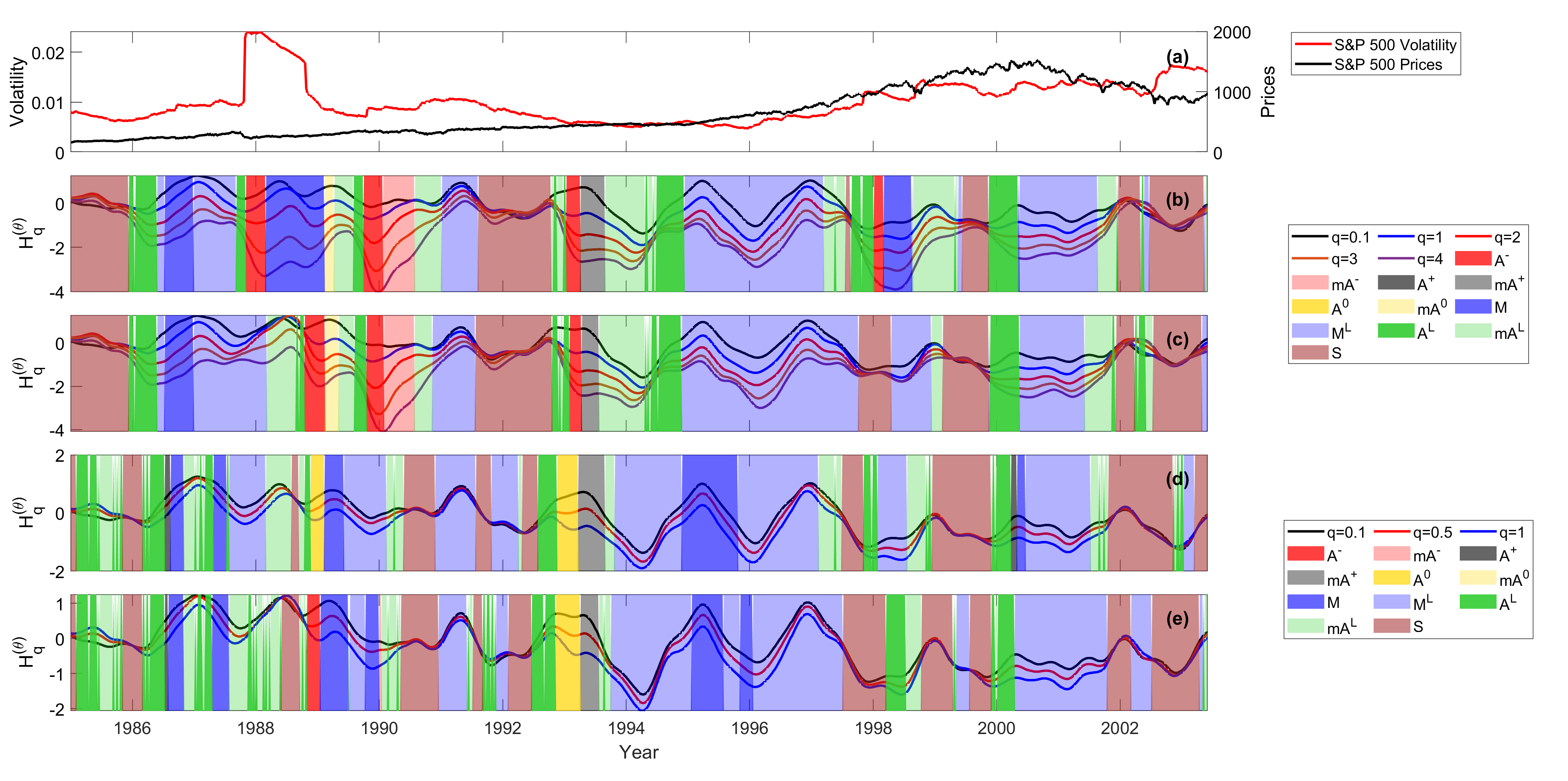}
\caption{\label{fig:SP500_zoom} SP500 index price time-series and scaling TPs in period 1985-2001: Comparison between TPs obtained from SP500 close prices (real index TP's) and TPs obtained after removing the 'black Monday', '1997 Asian crisis and '1998 Russian crisis' critical trading days (modified index TP's): (a) SP500 close prices, (b) real index $W_{14}$ TP's, (c) modified index $W_{14}$ TP's, (d) real index $B$-proxy TP's and (e) modified index $B$-proxy TP's. The 'warning' $A$-type TP's before 19th of October 1987 are maintained in the modified index results.  }
\end{center}
\end{figure}

\begin{figure}[h!]
\begin{center}
\includegraphics[height=0.6\textheight,width=1.05\textwidth]{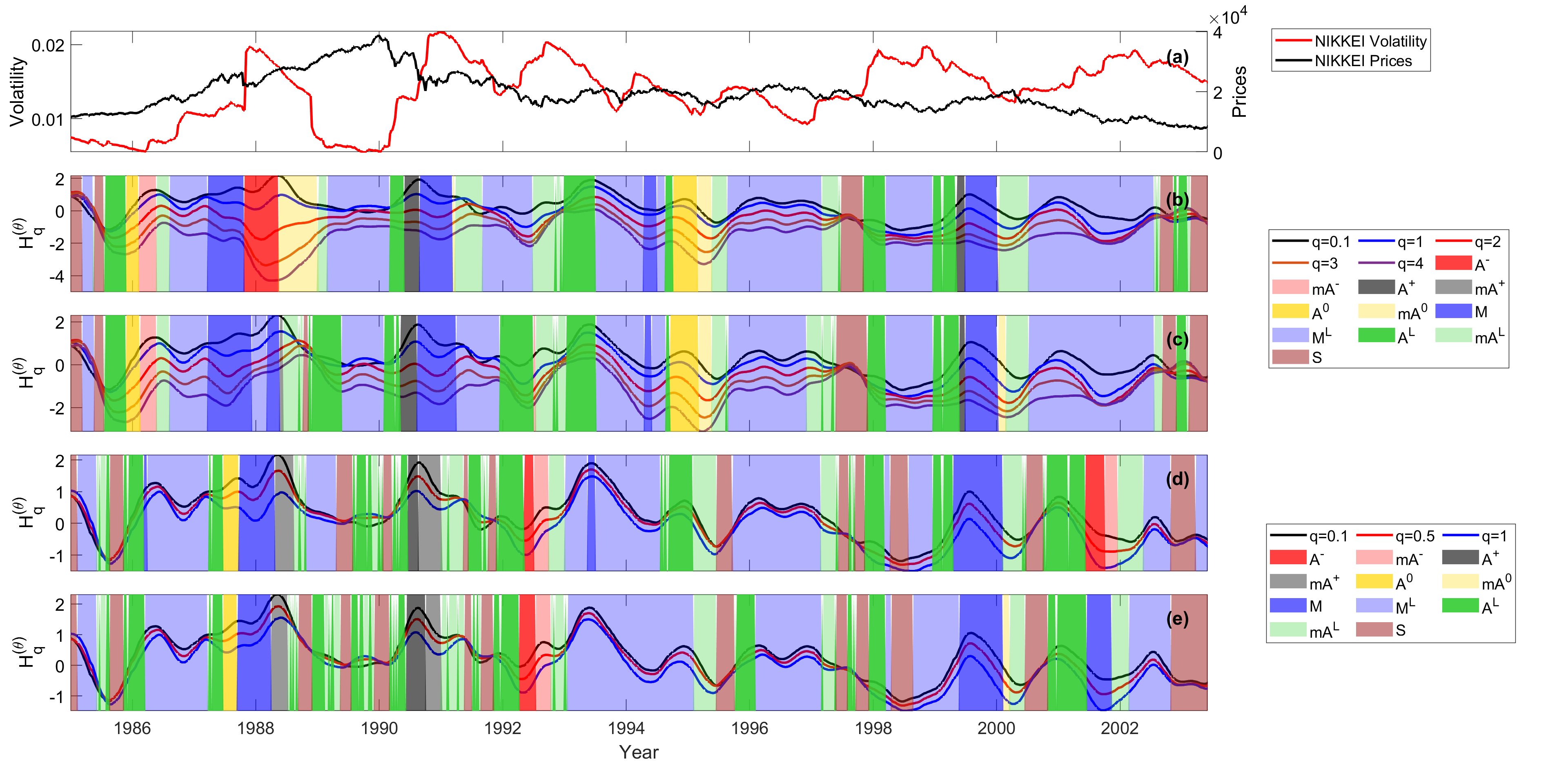}
\caption{\label{fig:NIKKEI_zoom} NIKKEI index price time-series and scaling TPs in period 1985-2001: Comparison between TPs obtained from NIKKEI close prices (real index TP's) and TPs obtained after removing the 'black Monday', '1997 Asian crisis and '1998 Russian crisis' critical trading days (modified index TP's): (a) Index close prices (b) real index TP's for $W_{14}$ (c) $W_{14}$ modified index TP's, (d) $B$-proxy real index TP's and (e) $B$-proxy modified index TP's.}
\end{center}
\end{figure}

\begin{figure}[h!]
\begin{center}
\includegraphics[height=0.6\textheight,width=1.05\textwidth]{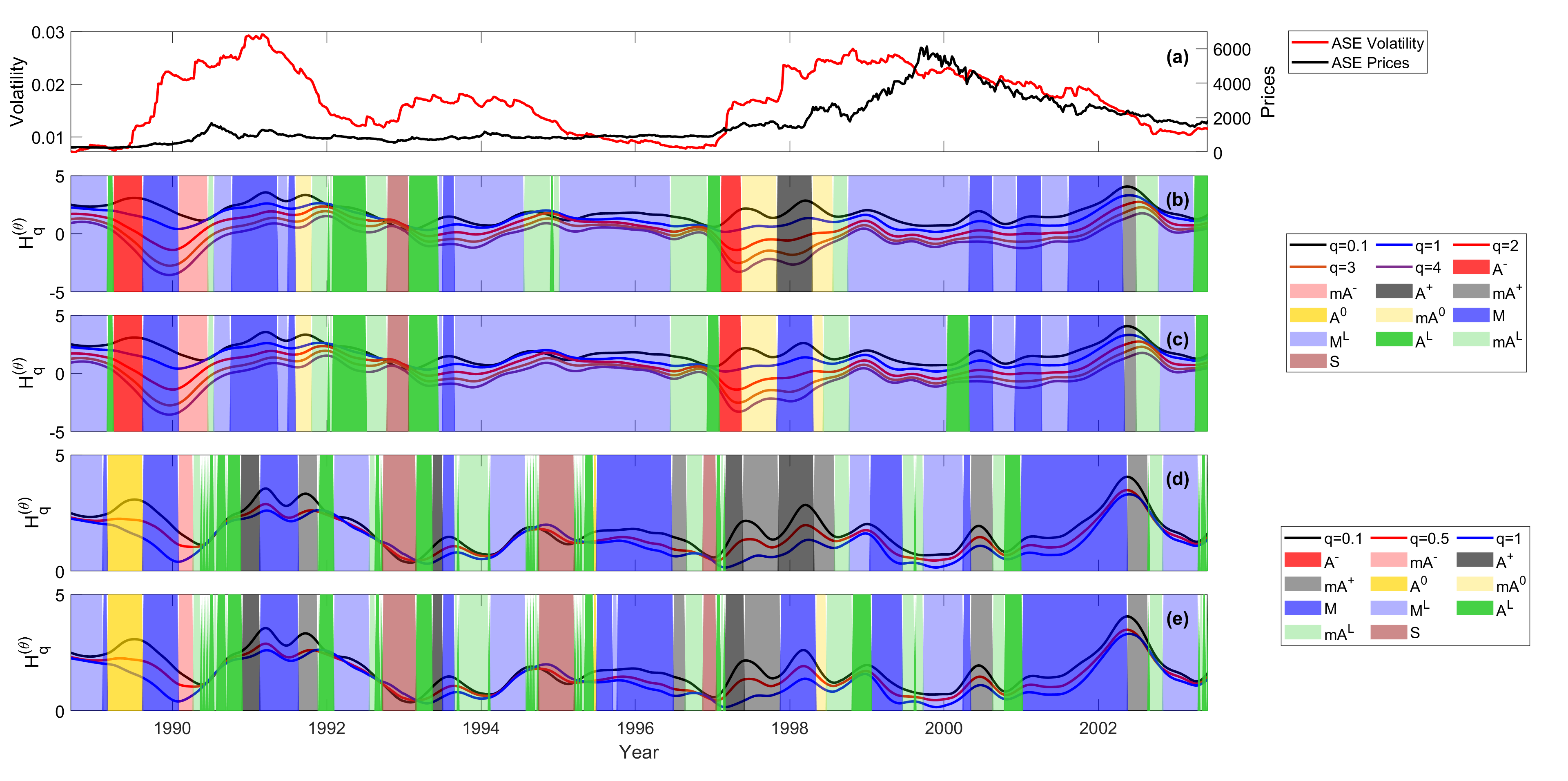}
\caption{\label{fig:ASE_zoom} ASE index price time-series and scaling TPs in period 1995-2001: As in figure \ref{fig:NIKKEI_zoom}, comparison between real index TP's and modified index TP's of ASE log close price $H_q$'s. The later are obtained after removing 1997 'Asian crisis' and 1998 'Russian crisis' critical trading days. (a), (b), (c), (d), (e) as in figure \ref{fig:NIKKEI_zoom}.}
\end{center}
\end{figure}

\begin{figure}[h!]
\begin{center}
\includegraphics[height=0.6\textheight,width=1.05\textwidth]{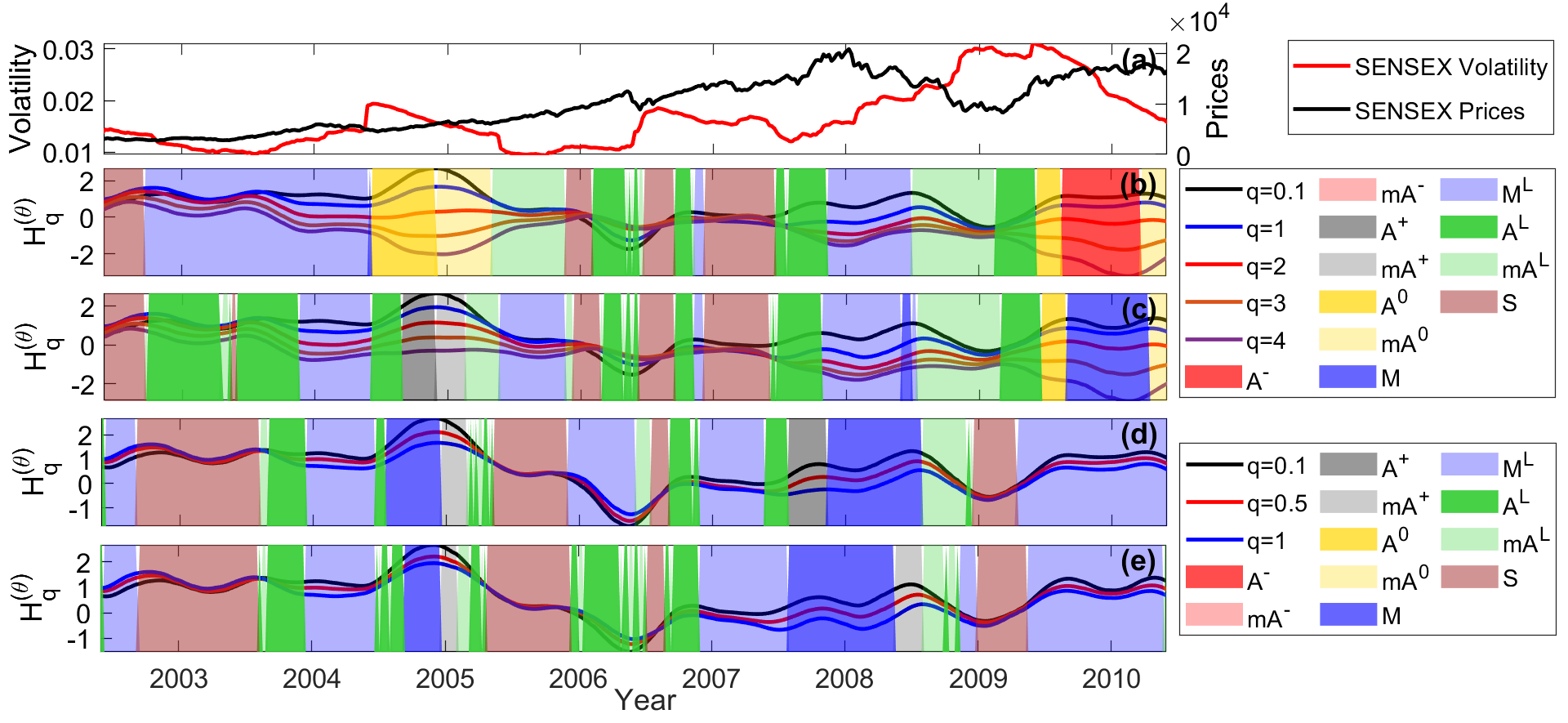}
\caption{\label{fig:SENSEX_zoom} SENSEX index price time-series and scaling TPs in period 1985-2001: Comparison between TPs obtained from SENSEX close prices (real index TP's) and TPs obtained after removing the 2006 'Ghandi election' crisis (modified index TP's): (a) S\&P~500 close prices (b) real index TP's for $W_{14}$ (c) $W_{14}$ modified index TP's, (d) $B$-proxy real index TP's and (e) $B$-proxy modified index TP's.}
\end{center}
\end{figure}

In figures~\ref{fig:SP500_zoom}~-~\ref{fig:SENSEX_zoom} a comparison between the \textit{w}GHE time series, $\gamma$ time-series and identified TP's of the real indices and the respective modified indices are shown, focusing on the time periods around the market events mentioned in Table~\ref{tab:DeletedDates}. For S\&P~500 and NIKKEI we observe that after B.M., the strong 'post-event' A$^-$ TP that exists in the real index data after Oct. 1987 (figures~\ref{fig:SP500_zoom},~\ref{fig:NIKKEI_zoom}b) and is directly related to the strong biasing induced to the tails of the price change distribution exclusively due to the four deleted days, is almost eliminated in the modified index $W'_{0.1,4}$ TP's (figures~\ref{fig:SP500_zoom},~\ref{fig:NIKKEI_zoom}c). However, the pre-event 'warning' A$^0$ TP and A$^L$ TP signals corresponding to a transition from a uniscaling to multiscaling starting in 1986, \textit{well before} the B.M. event are still present. Notice also, that an A$^L$ pattern well after the 'black Monday' event is still seen in the modified index TP's for NIKKEI, well before 1991. This suggests that the 'warning' $A$-type TP's observed well after Oct. 1987, are not an artefact of the B.M. event, but a consequence of market trading patterns before the NIKKEI 1991 bubble break-down. Similarly, the A$^-$ type TP before the year 2000 dot.com bubble is destroyed after in S\&P~500 after the deletion of the A.C. and R.C. related extreme tail events but there is a clear uniscaling to multiscaling transition via an A$^L$ TP well before the bubble break-down. The same A$^L$ TP is seen in NIKKEI, before 2000. Again, this suggests that the $A$-type warning signal in the period 1997-1998 is not exclusively an 'after-effect' product of the 1997 'Asian crisis' and 1998 'Russian crisis' events, but a product of trading dynamics of an extended period jut before the year 2000 bubble break-down.

The same conclusion can be drawn by looking at the 'warning' A$^-$ and A$^0$ types of TPs before the year 2000 ASE bubble in figure~\ref{fig:ASE_zoom}. These patterns are maintained, almost intact, in the modified index TPs, after removing A.C. and the R.C. related trading days from ASE. Again, this means that $A$-type patterns seen in the real index TPs are not just a product of 'after-effects' of one or two isolated big market events, but a product of an extended period of market trading patterns, well before a bubble bursts. This fact is particularly pronounced for the ASE 2000 bubble which was well under development in the period where the Asian and Russian crises occurred, since the the A$^-$ 'warning' TP's were barely affected by the removal of the few trading days related to these crises. However, the removal of the respective trading days before the break of the S\&P~500 'dot.com' bubble has had a different effect. We observe that the existing A$^-$ TP has almost disappeared and the dynamics approximately two years before the break of the bubble is uniscaling. However, even here, the dynamics undergoes a clear uniscaling to multiscaling transition almost a year before the bubble burst through and asymmetric A$^L$ TP. 

\section{Discussion}
\label{discussion}

By examining the GHE results for all these indices, we confirm that there are common elements among many of them especially in critical time periods, like a stock-market bubble or financial crisis. However, each index also has unique features indicating that the corresponding markets have different underling dynamics which can be related to global events for major stock indices and to local phenomena. In particular, we notice that critical events are usually driven by a uniscaling behaviour which is then followed by a usually sharp transition to multiscaling via 'asymmetric' multiscaling patterns. 

In some time periods one clearly sees that the $w$GHE's for higher values of $q$ show a sharp drop towards strongly anti-persistent behavior, whereas the respective small $q$ $w$GHE's are almost constant or rising, depicting neutral or persistent behavior. This behavior can be caused by two types of market changes: either (i) due to a critical single market event such as a market crash (e.g. Black Monday on Oct. 19, 1987 when S\&P~500 lost 20.4\% in one day), or (ii) a more extended critical period where the market behaves in a bullish way for small day-to-day price changes but shows anti-persistent behavior for large price changes (fig.~\ref{fig:SP500close}). In case (i), the large change is a single extreme tail point in the price change distributions taken within the particular time-window where the GHE's are calculated, a tail point that mostly affects the large $q$ GHE's causing a sharp drop. The volatility also shows a sharp rise at that date, followed by a gradual decay. The time duration of the effect of this single market event on the GHE time-series, as well as volatility, is in the order of $\Delta t$, the time window length used for calculating the price change distribution moments, and $\theta$ and leads to a pronounced A$^-$ pattern extended in time, although the actual scaling within this time window may be different. Subtracting this single event, would largely destroy the pattern, as it was seen, for instance, for the B.M. event in S\&P~500 and NIKKEI. In case (ii), we have shown that these patterns are a consequence of the increase of tail events that occur in a turbulent market period and the way they are correlated. For example, during a critical period (of a developing bubble, for example) there is increased frequency of large market drop tail events that are immediately followed (usually in the next trading day) by an equivalent rise. This combination of events occurring amidst a rising market trend, causes a sharp drop in the high-$q$ \textit{w}GHE's while the small $q$ \textit{w}GHE's are not so much affected. This type of market behavior that leads to market transition from a more 'regular' and efficient market (uniscaling behavior) to a more 'nervous' market has a plausible justification: when the majority of traders are afraid of or get the 'gut feeling' that the market becomes saturated and a crash is imminent, they are more likely to revert to rapid sales (in order to secure profits) that drive the market down by large amount during a single day. As the market is still in a rising trend, this sales spree is likely to be reversed and followed by a buying spree the next day in anticipation for a continued market rise. This sort of 'nervous' behavior was particularly notable with the ASE 2000 bubble, where the very pronounced asymmetric multiscaling patterns in the period 1997-1998 were not at all a result of just the Asian or Russian crises that took place within that period. In fact, one may argue, that even the large drops (followed by large rises) that are due to some justified market event (such as the A.C. and R.C. crises) are just a 'pretext' for a saturated market during a bubble development to correct itself. It is also a notable fact that the S\&P~500 changed from a rather long uniscaling (or very weakly multiscaling) period during the seventies to mid-eighties to a multiscaling period (transition is via an A$^L$ TP), starting in 1986 more than a year before Black Monday, a clear stock-market historical event that remains completely unexplained by the economic surroundings of the preceding period. In a large survey carried out by Shiller \cite{Shil1988} over a sample of more than 800 investors, when interviewed, the most frequent answer they gave to the question why they behaved the way they did during that day, was that they had a 'gut-feeling' that there was an impeding crash. This 'gut-feeling' was captured by \textit{w}GHE's measuring the market scaling transition that occurred long before the event, as these traders developed particular trading habits which, over a rather long period before the crash, lead to a sequence of tail events that would spark an A$^L$ TP and multiscaling behavior seen in 1986-1987.  

In conclusion, case (i) 'A' patterns caused by large single events can be distinguished from extended time (case (ii)) 'A' patterns, by the fact that the first \textit{follow} a crash or a bubble-break, whereas the second are \textit{preceding} a possible crash or bubble-break. In this sense, 'A' patterns, especially when they follow a period of uniscaling behavior, can be used as warning signals for critical market time periods.

We also noticed some differences between major and peripheral markets. In particular, in major stock indices, more abrupt transitions between patterns are observed during critical time periods, while for peripheral markets they are much smoother. This is probably due to the number of market participants and the amount of information available to them. In fact, in a global market the market shift due to 'bad news' can completely alter market dynamics in a relatively small amount of time. A second difference lies in the fact that for major indices, multiscaling is not associated directly to period of a of recession or crisis while it is mostly the case for peripheral markets. This is probably due to the fact that in peripheral markets there isn't enough liquidity to absorb the huge heterogeneity generated by the market participants.

\section{Conclusions}
\label{conclusions}

In this paper, we have presented for the first time how different temporal patterns can emerge from the dynamics of the time-dependent generalized Hurst exponents (GHE). In particular, we proposed several patterns which differentiate uniscaling from multiscaling and further differentiate two forms of multiscaling, i.e. symmetric and asymmetric multiscaling in the temporal evolution of GHE timeseries. These temporal patterns combined with the analysis of the multiscaling width $W$ and the multiscaling depth $B$ (and their dynamics) offer an important set of tools to signal critical events in financial time series and not only. We also introduced a completely algorithmic and general procedure to identify such patterns in any time-series of GHE's, which allows one to determine these patterns in a statistically significant manner. Regarding the calculation of the GHE time-series, we also addressed the important issue of choosing a proper sliding time window length $\Delta t$ and provided an empirical rule that is based on minimising the noise due to finite-size effects in the GHE calculations and at the same time capturing the actual local dynamical changes of the scaling over short time scales. 

Results showed very interesting patterns among major and peripheral markets. We found similar patterns among the market considered but also differences related to local behaviors. One of the common features is the existence of a (usually sharp) transition from an 
uniscaling to multiscaling pattern in the rising period, before a stock market bubble breaks, such as the 2000 bubble in S\&P~500 and ASE or the 1991 bubble for NIKKEI. ASE, being a small and peripheral market with low liquidity, had a much more pronounced and robust 'asymmetric' multiscaling warning pattern before its large 2000 bubble, than major indices like S\&P~500 and NIKKEI. This feature is also present in stock market crises that are externally caused, such as the 2008 real-estate market crisis, but in a significantly weaker form. For example, for ASE, whereas the 2000 and 1990 crashes showed very pronounced and clear A$^-$ patterns, the 2008 crash showed a weaker A$^L$ pattern. Another feature is that there exists some kind of notable scaling transition shortly before or after the break-down of a bubble, usually a change from a strong asymmetric multiscaling to either an uniscaling or moderate multiscaling TP. It should be stressed that the transition to an asymmetric multiscaling TP is manifested by past data sufficiently prior to the bubble breakdown so that this feature could be used as a 'warning' signal of a bubble in development, in particular, if the strong multiscaling is accompanied to relatively low (and maybe rising) volatility. In general, transitions always occur at some critical date when there is either the beginning of a new period of development or the end of some type of crisis. However, if we are talking about a global crisis, various stock markets are affected in significantly different ways. The differences are pronounced if major indices, that highly correlate to global events are compared to peripheral or developing markets which are mostly affected by local events. Indeed, major market crashes also affect the scaling behaviour of peripheral markets, while the reverse is not true. For several indices there are extended time periods of uniscaling behavior and time periods of clear multiscaling behavior, while some indices are, on the overall, more multiscaling than others. The rich variety of information that can be conveyed by the newly introduced scaling patterns can be used as a valuable tool to obtain the 'fingerprint' of a possible turbulent market period and also issue warning signals for impeding market crashes or other critical events.

Finally, as the defined scaling temporal patterns are clearly related to the details of the underlying complex dynamics of the physical system in a physically justifiable way, they offer (together with the algorithmic identification procedure presented in this work) a tool to characterise the dynamical evolution of scaling of any complex system. Of particular interest would be to apply the temporal pattern analysis presented in this work to low-dimensional complex systems that contain, apart from fat-tailed change difference distributions, enhanced temporal correlations. These systems would be optimal test-beds for testing the 'predicting power' of GHE's for the future evolution of the system, especially for critical events. Future work will be devoted to disentangling the effect of fat tails and correlation to the various forms of multiscaling in a robust statistical manner. A second extension to this work will be in the direction of the \textit{quantification} of the asymmetries (based on empirically defined metrics that would depend on the GHE temporal profiles) in order to algorithmically detect strong asymmetries in scaling which can be used by market participants for trading strategies, \textit{e.g.} issue a sell order when the asymmetry is higher than the long term asymmetry during specific market conditions such as a bullish market. The construction of a total 'market risk' indicator that could depend on these GHE metrics is another very interesting possibility for future work. Such an indicator would be a very useful tool in the hands of investors and policy makers in order to detect and quantifiably assess financial risk.

\bibliography{Paper_Hurst_tool}
\bibliographystyle{elsarticle-num-names} 

\section{Appendix}
\label{appendix}

\begin{figure}[h!]
\begin{center}
\includegraphics[height=0.6\textheight,width=1.05\textwidth]{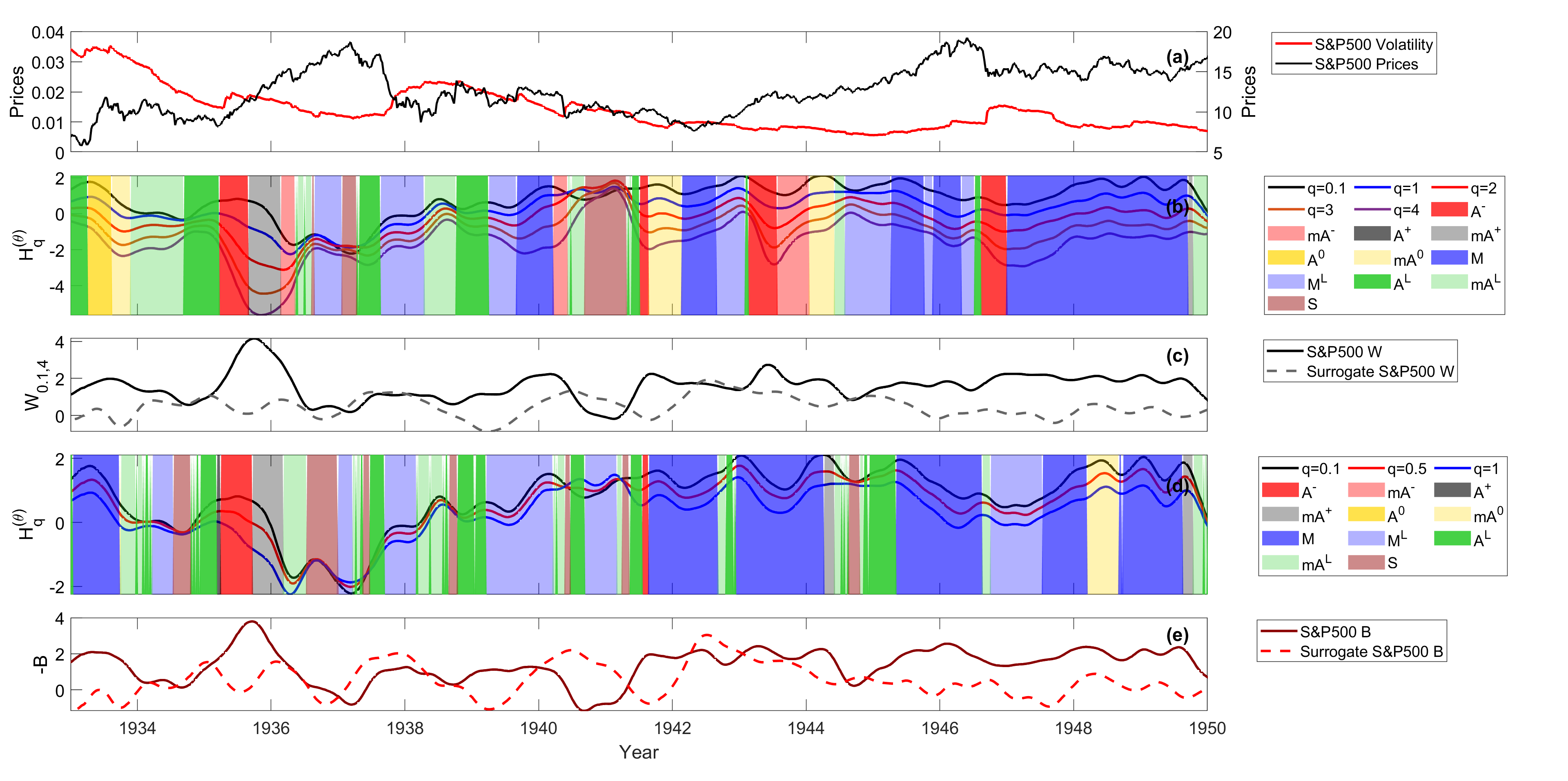}
\caption{\label{fig:SP500_zoom33-50} S\&P~500 blow-up of period 1933-1950: (a),(b),(c),(d) and (e) exactly as described in caption of figure~\ref{fig:SP500close}.}
\end{center}
\end{figure}

\begin{figure}[h!]
\begin{center}
\includegraphics[height=0.6\textheight,width=1.05\textwidth]{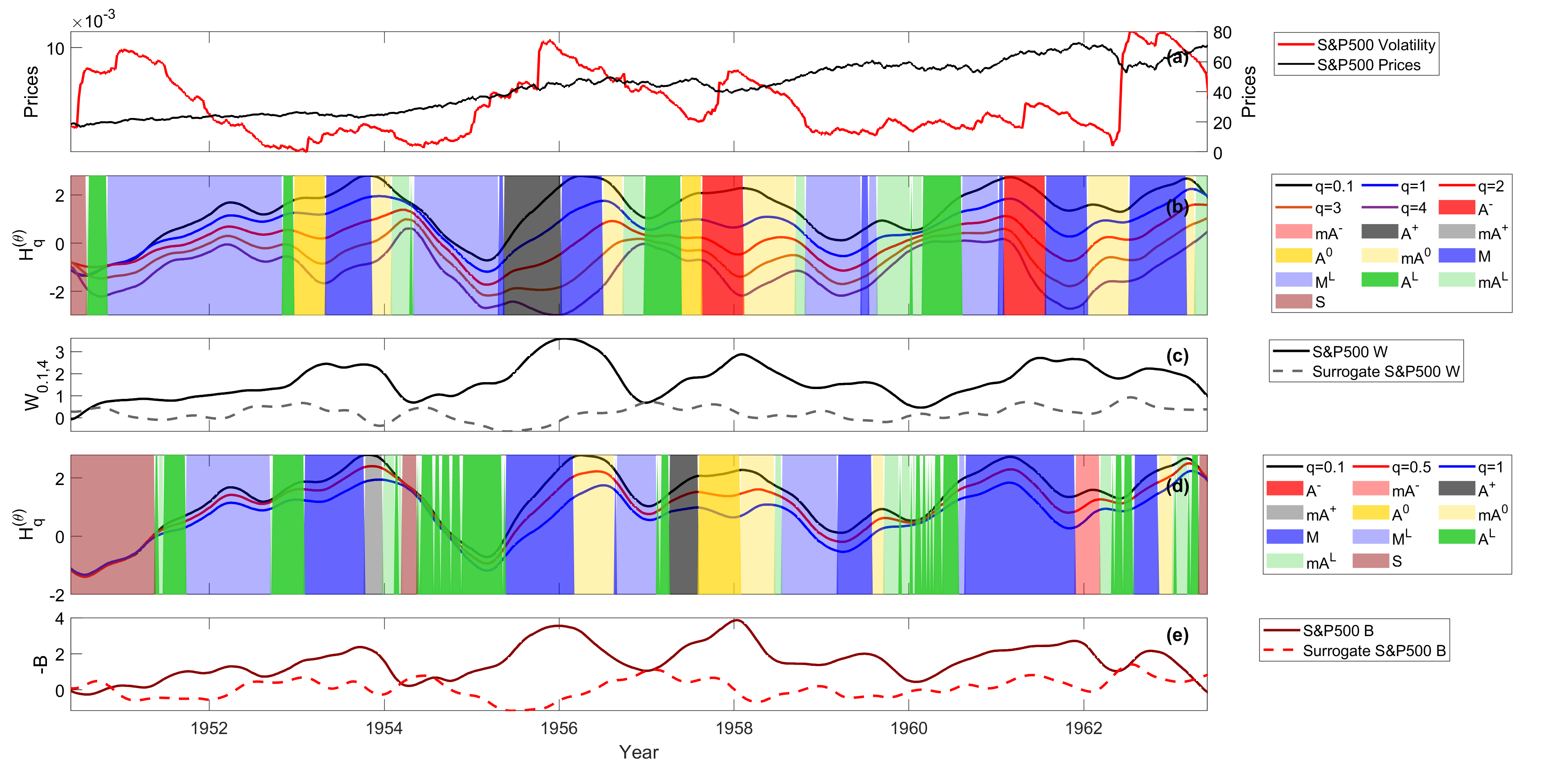}
\caption{\label{fig:SP500_zoom50-63} S\&P~500 blow-up of period 1950-1963: (a),(b),(c),(d) and (e) exactly as described in caption of figure~\ref{fig:SP500close}.}
\end{center}
\end{figure}

\begin{figure}[h!]
\begin{center}
\includegraphics[height=0.6\textheight,width=1.05\textwidth]{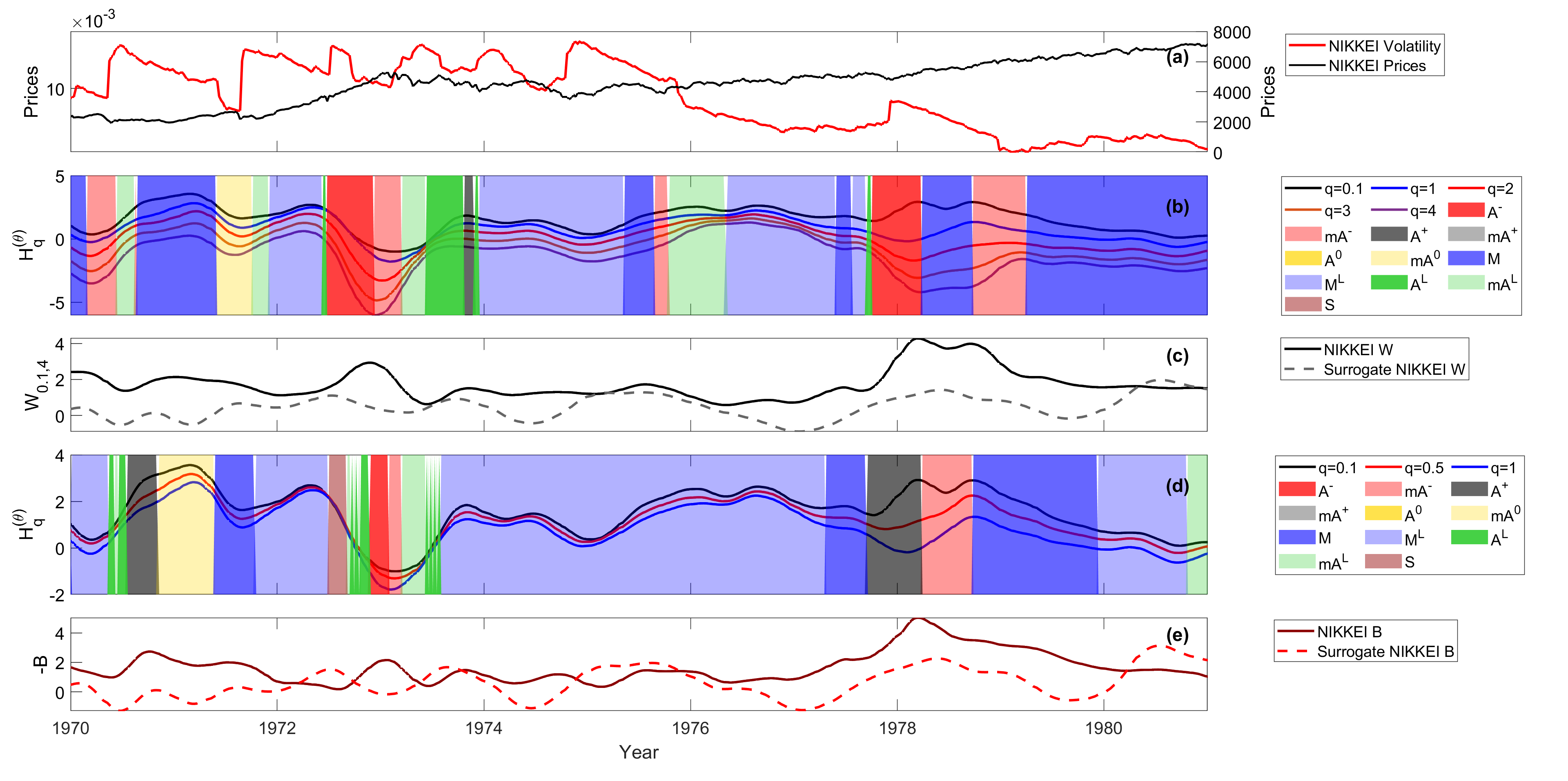}
\caption{\label{fig:NIKKEI_zoom70-80} NIKKEI blow-up of period 1970-1980: (a),(b),(c),(d) and (e) exactly as described in caption of figure~\ref{fig:SP500close}.}
\end{center}
\end{figure}

\end{document}